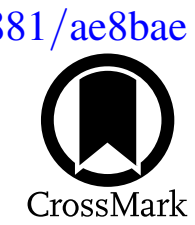

# A Streamer Driving Misalignment in the Circumtriple Disk of GW Ori

Maria Galloway-Sprietsma[1], Jaehan Bae[1], Toni Phillips[1], Jane Huang[2], Myriam Benisty[3], Matthew Porter[1], Christian Ginski[4], and Andrew Winter[5]

[1] Department of Astronomy, University of Florida, Gainesville, FL 32611, USA
[2] Department of Astronomy, Columbia University, 538 W. 120th St., Pupin Hall, New York, NY 10027, USA
[3] Max-Planck Institute for Astronomy (MPIA), Königstuhl 17, 69117 Heidelberg, Germany
[4] School of Natural Sciences, Center for Astronomy, University of Galway, Galway H91 CF50, Ireland
[5] Queen Mary University of London, Mile End Rd., E14NS, London, UK

## Abstract

GW Ori is a triple stellar system with a disk featuring three misaligned dust rings. We present Atacama Large Millimeter/submillimeter Array molecular line data of CO isotopologues, revealing a streamer feeding the disk in $^{12}$CO and $^{13}$CO. Through multipointing observations, we find the streamer extends to ∼30″ on-sky (∼12,000 au) and estimate a mass of 1.6 $M_{\rm Jup}$ using $^{13}$CO. Fitting the morphology and kinematics of the streamer, we estimate a mass infall timescale of 0.04 Myr and a mass infall rate of $3.6 \times 10^{-8}\ M_\odot\ {\rm yr}^{-1}$, approximately an order of magnitude lower than the stellar accretion rate. Our best-fit trajectory shows that the streamer meets the disk at the outermost dust ring and that the streamer's angular momentum vector is closely aligned with this outermost ring within ≈3°, in contrast to the ≈32° misalignment with the innermost ring, suggesting the streamer is the likely cause of the misalignment. However, the total angular momentum of the streamer is smaller than that of GW Ori's disk; this, together with the low accretion rate, indicates that we are probably witnessing the end stages of infall. Total Power observations reveal bright emission connecting the streamer to the surrounding star-forming region, with the projected distance falling well within the Bondi–Hoyle radius, implying the streamer could have originated through Bondi–Hoyle accretion as GW Ori moves through its natal cloud. Together, these results suggest that GW Ori remains dynamically linked to its parental cloud through ongoing accretion, with the streamer serving as the likely driver of its misaligned disk structure.

## 1. Introduction

Stars and their planetary systems form from interstellar material over millions of years. Classical star formation models consider collapsing clouds as isolated systems, where the material in the cloud is accreted onto the star and the disk in a constant and axisymmetric manner (F. H. Shu 1977). In reality, star-forming regions are far more complex, shaped by turbulent flows, dynamical interactions, and asymmetries that drive nonuniform patterns of collapse and accretion (e.g., M. R. Bate et al. 2003; C. J. Lada & E. A. Lada 2003; J. Ballesteros-Paredes et al. 2007; C. F. McKee & E. C. Ostriker 2007).

In recent years, observations of protoplanetary disks have shown evidence of interaction between the parental molecular cloud and the disk in the form of infalling material. These so-called "streamers" have been observed around disks of all evolutionary stages, including both embedded Class 0/I disks (J. J. Tobin et al. 2012; S. Takakuwa et al. 2013; H.-W. Yen et al. 2014; J. E. Pineda et al. 2020; N. M. Murillo et al. 2022; T. J. Thieme et al. 2022; T.-H. Hsieh et al. 2023; J.-E. Lee et al. 2023; S. Mercimek et al. 2023; T. Hanawa et al. 2024; M. T. Valdivia-Mena et al. 2024; X. Mai et al. 2025; F. A. Olguin et al. 2025) and more evolved Class I/II disks (E. Akiyama et al. 2019; H.-W. Yen et al. 2019; F. O. Alves et al. 2020; D. Mesa et al. 2022; A. Garufi et al. 2024; A. Gupta et al. 2026; M. Tanious et al. 2025); see also A. Gupta et al. (2023) and J. E. Pineda et al. (2023) for a review. The growing number of observations indicates that anisotropic infall could be commonplace, and may play a critical role in the evolution of disks and the subsequent planet formation therein. Simulations suggest that infall can produce pressure bumps where vortices can form through the Rossby wave instability (RWI; J. Bae et al. 2015; A. Kuznetsova et al. 2022). Infall through streamers may also misalign disks (C. Ginski et al. 2021; M. Kuffmeier et al. 2021; M. Tanious et al. 2025), induce gravitational instability (K. M. Kratter et al. 2008; J. Bae et al. 2014; J. Speedie et al. 2025), and drive angular momentum transport (G. Lesur et al. 2015). In turn, streamers can directly influence the planet formation process, and can have broad impacts on the dynamics of the protoplanetary disks accreting them. Stellar multiplicity may further enhance the likelihood of infall, as multiple systems are less likely to have undergone a gravitational kick away from the initial overdensity, and are therefore more likely to be at the location where filament flows converge (M. R. Bate et al. 2003; M. Kuffmeier et al. 2023). A key question, however, is whether observed streamers possess the physical and kinematic properties predicted by models. Establishing the observational–theoretical connection is essential for determining their true impact on disk evolution.

**Table 1**
Summary of the ALMA Observations Used to Image GW Ori and the Associated Streamer in This Work

| Project ID | Array | Pointing | ΔR.A. | ΔDecl. | Angular Res. (arcsec) | Integration Time (minutes) | Obs. Date | # of EBs | Max. Reco. Scale (arcsec) |
|---|---|---|---|---|---|---|---|---|---|
| 2017.1.00286.S | 12 m C-7 | Center | 0 | 0 | 0.116 | 94 | 2017-12-10 | 1 | 2.3 |
| 2017.1.00286.S | 12 m C-3 | Center | 0 | 0 | 0.730 | 38 | 2018-04 | 2 | 9.0 |
| 2022.1.01108.S | 12 m C-3 | Offset | $0^{\mathrm{s}}.76$ | $6''.5$ | 0.733 | 31 | 2022-12-27 | 1 | 8.7 |
| 2022.1.01108.S | 7 m | Center | 0 | 0 | 5.197 | 24 | 2023-01-23 | 1 | 31.6 |
| 2022.1.01108.S | 7 m | Offset | $0^{\mathrm{s}}.76$ | $6''.5$ | 5.635 | 74 | 2023-05-19 | 1 | 31.1 |
| 2022.1.01108.S | TP | Center | 0 | 0 | 23.179 | 52 | 2023-05 | 2 | 410.9 |
| 2022.1.01108.S | TP | Offset | $0^{\mathrm{s}}.76$ | $6''.5$ | 23.179 | 231 | 2023-05 | 4 | 410.9 |

**Notes.** The center pointings are toward the central star at R.A. = $5^{\mathrm{h}}29^{\mathrm{m}}7^{\mathrm{s}}.988$ and decl. = $+11°52'14''.5$. The offset pointings are $1/2$ the half-power of the 12 m beamwidth from the central star, at R.A. = $5^{\mathrm{h}}29^{\mathrm{m}}7^{\mathrm{s}}.22$ and decl. = $+11°52'21''$. For the observations with multiple execution blocks, we list just the month and year that the observations were taken.

Here, we focus on the Class II source GW Ori, for which previous observations have revealed evidence of late-stage infall. GW Ori is a triple stellar system (J. P. Berger et al. 2011) in $\lambda$ Ori, approximately 402 $\pm$ 10 pc away (Gaia Collaboration et al. 2018). The primary star, GW Ori A, is a G8 pre-main-sequence star, with companion GW Ori B at $\sim$1 au separation and the third companion GW Ori C at an $\sim$8 au separation (J. P. Berger et al. 2011; M. Fang et al. 2014). The age of the system is estimated to be 0.3–1.3 Myr (I. Czekala et al. 2017). Surrounding the system is a massive ($\sim$0.12 $M_{\odot}$, out to 1300 au; M. Fang et al. 2017) circumtriple disk, first inferred from the spectral energy distribution (SED; R. D. Mathieu et al. 1991) and later confirmed through millimeter continuum and molecular line observations (R. D. Mathieu et al. 1995; M. Fang et al. 2014, 2017). Based on the SED, GW Ori is classified as a Class II source.

Extensive observations targeting both the continuum and bright CO isotopologues ($^{12}$CO, $^{13}$CO, and C$^{18}$O) have been used to constrain the stellar masses and disk structure. Kinematic modeling of CO emission yields stellar masses of $M_{\rm A} \approx 2.7\ M_{\odot}$, $M_{\rm B} \approx 1.7\ M_{\odot}$, and $M_{\rm C} \approx 0.9\ M_{\odot}$ (I. Czekala et al. 2017), while orbital modeling produces comparable values ($M_{\rm A} \approx 2.47\ M_{\odot}$, $M_{\rm B} \approx 1.43\ M_{\odot}$, and $M_{\rm C} \approx 1.36\ M_{\odot}$; S. Kraus et al. 2020), implying a total system mass of $\sim$5.3 $M_{\odot}$. Previous Atacama Large Millimeter/submillimeter Array (ALMA) continuum observations revealed three dust rings, with the inner two centered on the A–B binary, and an outer one which is offset (J. Bi et al. 2020). The rings are misaligned with respect to the orbital plane of GW Ori A and B, with inclinations of ~11° (inner ring), ~35° (middle ring), and ~40° (outer ring) (J. Bi et al. 2020; see their Figures 2 and 3 of S. Kraus et al. 2020). Previous studies have attempted to explain this misalignment through the gravitational torques of the stellar companions (S. Kraus et al. 2020) or a yet-unseen planetary companion in the disk (J. L. Smallwood et al. 2021). In this work, we consider an alternative scenario in which the misalignment is instead driven by late-stage infall.

Previous SMA observations have shown that GW Ori has a streamer visible on the northwestern side of the disk in both $^{12}$CO and $^{13}$CO, first noted in M. Fang et al. (2017). Extended emission in IR data is also seen beyond the continuum rings (S. Kraus et al. 2020). The presence of this large-scale structure suggests that GW Ori is still accreting material from its surrounding cloud, despite its advanced evolutionary stage. Continued accretion such as this may play a critical role in shaping the disk's unique morphology, kinematics, and misalignment, potentially influencing ongoing planet formation within the system. Altogether, the GW Ori disk provides a prime opportunity to study how complex stellar environments shape disk structure and evolution.

In this work, we study the nature of this infall to understand how it impacts the disk by combining new and archival ALMA observations. Section 2 describes the ALMA observations used in this study. Section 3 describes the techniques used to derive the streamer properties. We demonstrate that the angular momentum vector of the streamer is closely aligned with that of the outer dust ring, providing observational evidence that the streamer is the source of the disk misalignment. We further show that the current infall rate is about an order of magnitude below the stellar accretion rate, suggesting that we are witnessing the end stages of this infall event. We present a discussion of the findings in Section 4, and the conclusions are summarized in Section 5.

## 2. Observations

This analysis utilized ALMA data from programs 2017.1.00286.S (PI: Muto) and 2022.1.01108.S (PI: Galloway-Sprietsma). Below, we discuss both the archival and the new data, and how they were combined and imaged.

### 2.1. Archival Data

ALMA program 2017.1.00286.S observed GW Ori in ALMA Band 6 using the 12 m array in the C-7 and C-3 configurations with a minimum velocity resolution of 0.092 km s$^{-1}$ and angular resolutions of 0.″116 and 0.″730, respectively. The spectral setup targeted CO isotopologues in Band 6: $^{12}$CO $J = 2-1$, $^{13}$CO $J = 2-1$, and C$^{18}$O $J = 2-1$ at rest frequencies of 230.538 GHz, 220.399 GHz, and 219.55 GHz. The streamer is visible in both $^{12}$CO and $^{13}$CO, but is cut off at the edges due to the limited field of view (FOV) of 25″ and spatial sampling. Therefore, we requested additional observations offset from the central disk to image the full extent of the streamer.

### 2.2. New Data

Program 2022.1.01108.S was specifically designed to characterize the entire extent of three disks with streamers to probe their mass and kinematics. For this purpose, this program utilized multiple pointings from all three ALMA arrays (12 m, 7 m, and Total Power) in order to fill in the visibility space and probe a broad range of spatial scales. Toward GW Ori, we obtained a set of offset pointings with the 12 m, 7 m, and Total Power arrays, and a set of center pointings with the 7 m and Total Power arrays. The observations used to image GW Ori and the associated streamer are listed in Table 1. The spectral setup, the same as for program 2017.1.00286.S, was comprised of four spectral windows (`spw`) in Band 6 to probe $^{12}$CO $J = 2-1$ (`spw` 1), $^{13}$CO $J = 2-1$ (`spw` 2), C$^{18}$O $J = 2-1$, SO $J = 6-5$ (`spw` 3), and continuum (`spw` 4) at rest frequencies of 230.538 GHz, 220.399 GHz, 219.94 GHz, 235.141 GHz, and 233 GHz, respectively. Spectral windows 1 and 2 had a spectral sampling of 30.519 kHz = 0.04 km s$^{-1}$, `spw` 3 had a spectral sampling of 61.035 kHz = 0.083 km s$^{-1}$, and all three had a total bandwidth of 58.59 MHz = 80 km s$^{-1}$. The fourth `spw`, reserved for continuum observations, had a bandwidth of 1875.0 MHz = 2413 km s$^{-1}$. Longer integration times were chosen for the offset pointings as we expected the streamer density to drop as a function of distance; these times are shown in Table 1. Varying angular resolutions were achieved with the three antenna arrays: 0.″733 with the 12 m array, 5.″635 with the 7 m array, and 23.″179 with the Total Power array.

### 2.3. Imaging

To explore the full extent of the streamer, the new and archival observations were combined. All new data was reduced using CASA version 6.4.0 (CASA Team et al. 2022) after the ALMA pipeline calibration was performed. The archival data and the 7 m central and offset pointings were self-calibrated; the new 12 m pointing did not include continuum emission and therefore was not self-calibrated.

The multiscale CLEAN algorithm (T. J. Cornwell 2008) was used to image the data, implemented via tclean in CASA. We used a robust parameter of 0.5, a UV tapering of 1.″0, and a cycle factor of 2. We utilized the "mosaic" function within tclean to image the multiple 12 m pointings from programs 2017.1.00286.S and 2022.1.01108.S, combining the archival data with the offset pointings. The same methodology was used to image the 7 m data and the 12 m + 7 m data. From this, we obtained three image sets: High Resolution Images (which use the 12 m extended configuration (C-7) and the compact configuration (C-3) central and offset pointings), Low Resolution Images (7 m central and offset pointings), and Combined Images (12 m TM1 and TM2 plus 7 m central and offset pointings). We achieved rms values of 0.0039 Jy beam$^{-1}$ ($^{12}$CO), 0.0020 Jy beam$^{-1}$ ($^{13}$CO), and 0.0014 Jy beam$^{-1}$ (C$^{18}$O) for the High Resolution Images; 0.077 Jy beam$^{-1}$ ($^{12}$CO), 0.030 Jy beam$^{-1}$ ($^{13}$CO), and 0.019 Jy beam$^{-1}$ (C$^{18}$O) for the Low Resolution Images; and 0.0064 Jy beam$^{-1}$ ($^{12}$CO), 0.0034 Jy beam$^{-1}$ ($^{13}$CO), and 0.0023 Jy beam$^{-1}$ (C$^{18}$O) for the Combined Images.

The Total Power data was imaged using CASA's single-dish imaging pipeline, tsdimaging, which uses an on-the-fly mapping technique (J. G. Mangum et al. 2007). To combine Combined Images with the Total Power data, we used CASA's feather algorithm, which combines two images using their Fourier transforms (see S. Stanimirovic 2002; T. T. Helfer et al. 2003; J. Koda et al. 2011). We determined the single-dish image flux scaling factor, sdfactor, by taking the flux ratio between the interferometric (IF) image and the Total Power image. We found a flux ratio of 1.26, and used this as the sdfactor; this value is consistent with other works (A. Plunkett et al. 2023). This feathering process results in the Combined and Feathered Images, which utilize all arrays for the maximum amount of uv-coverage. The bulk analysis in this paper uses the Combined and Feathered Images of $^{12}$CO and $^{13}$CO, which are a combination of the 12 m, 7 m, and Total Power visibilities. We achieved rms values of 0.0068 Jy beam$^{-1}$ ($^{12}$CO), 0.0035 Jy beam$^{-1}$ ($^{13}$CO), and 0.0023 Jy beam$^{-1}$ (C$^{18}$O) for the Combined and Feathered Images.

## 3. Results

### *3.1. Moment and Channel Maps*

We present peak intensity (moment-8), integrated intensity (moment-0), and velocity (moment-1) maps of the Combined and Feathered images for $^{12}$CO, $^{13}$CO, and C$^{18}$O $J = 2\text{–}1$ in Figure 1. The same moment maps for the low-resolution 7 m data cubes are shown in Appendix A. The intensities are converted to brightness temperature using the full Planck law. In $^{12}$CO, the archival observations reveal a streamer extending ∼14″, or ∼5600 au, from the triple system. With the addition of the new data, the streamer can be seen beyond this. The peak intensity map (leftmost panel, Figure 1) showcases the disk and the full extent of the streamer.

The Keplerian disk is centered at (0″, 0″), with an approximate radius of 3″ (∼1200 au). Past the disk, the main component of the infalling material in $^{12}$CO, seen to the northwest (NW), reaches out to ∼30″ (∼12,000″) on-sky. Additional material can be seen surrounding the disk, particularly to the north, with small amounts extending to the east of the disk as well. It is unclear if this material is associated with the streamer, but it appears to be related to the local emission as the velocities are close to that of the disk. The streamer is also clearly visible in the moment-8 map of $^{13}$CO, extending again to the northwest from the central disk. The material reaches out to ∼ 18″ (∼7200″) on-sky, and is aligned with the $^{12}$CO emission. Only the central disk was detected in C$^{18}$O, which can be seen in the bottom panel of Figure 1.

The bulk of the $^{12}$CO streamer emission has peak brightness temperatures of 20 K, implying that it is optically thick. The temperature of the streamer drops along the edges, and it is likely that the outside edges of the streamer are at least moderately optically thin. Most of the $^{13}$CO emission is likely optically thin, evidenced by the colder measured brightness temperatures within the streamer (<7 K). An area of reduced intensity can be seen through the center of the streamer in both $^{12}$CO and $^{13}$CO, particularly visible in the peak intensity map of $^{13}$CO (see Figure 1). This could be due to an optical depth effect, a rotation, or it is possible that two separate streamers are falling onto the GW Ori disk. These possibilities will be explored in more detail in Sections 3.4 and 4.2.

The velocity (moment-1) maps of $^{12}$CO and $^{13}$CO are shown in the rightmost panels of Figure 1. The Keplerian disk can be seen clearly in $^{12}$CO and $^{13}$CO, with the northern and southern sides of the disk redshifted and blueshifted, respectively. In both molecular lines, the bulk of the streamer is blueshifted with respect to the systemic velocity (13.5 km s$^{-1}$), but the emission that extends beyond 15″ gradually changes from blueshifted to redshifted over its length. We discuss the large-scale kinematics of the streamer in Section 4.3. The double structure seen in the peak intensity maps can also be seen in the velocity maps, particularly in $^{13}$CO. The southernmost component of the structure moves at slightly faster speeds than the line-of-sight (LOS) speed and the northernmost component. This unique morphology will be discussed in Section 4.2.

Figure 2 showcases eight channel maps that span some of the velocity range of the infalling material in $^{12}$CO. The associated channel maps for $^{13}$CO are overlaid in cyan, with contours at 4 K. The bulk of the streamer is clearly visible at 13.7 km s$^{-1}$. At this velocity, there is additional emission toward the north and west of the disk. As we decrease in velocity (blueshift), the streamer comes closer to the disk and appears to fully connect at about 13.1 km s$^{-1}$. The material to the north of the disk persists throughout these channels, and may be tied to the infall interaction or surrounding medium. The streamer in $^{13}$CO closely follows the $^{12}$CO morphology, although it does not extend out as far and is centralized to the brightest parts of the $^{12}$CO emission. The $^{13}$CO emission is optically thin (as its brightness temperatures are low), and therefore allows us to trace regions deeper within the infalling material, whereas $^{12}$CO traces the surface of the infall. We show the complete channel maps that include all channels with streamer emission in Figures 13 and 14 in Appendix A.

We did not detect SO toward the disk or the streamer with our observational setup. SO has been identified as a reliable shock tracer in systems with streamers (e.g., H.-W. Yen et al. 2014; A. Garufi et al. 2022; M. Tanious et al. 2024). However, the nondetection of SO in this work does not necessarily imply a lack of shocks, but is likely related to our achieved sensitivity and signal-to-noise ratio (SNR). Sulfur-bearing species such as SO are tentatively detected in other archival ALMA data of GW Ori (2021.1.01561.S, 2017.1.00286.S) but we leave this analysis for future work.

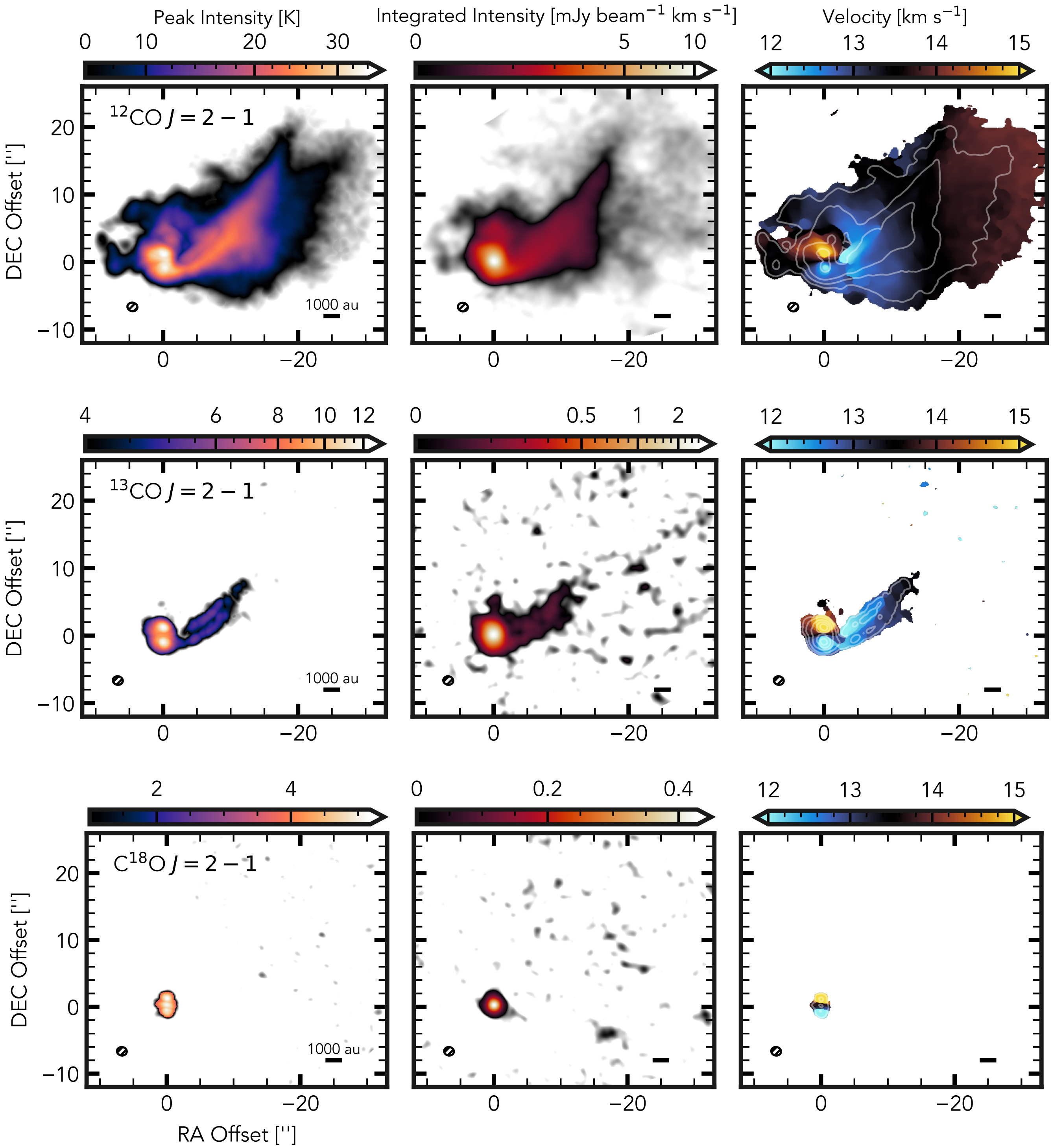

**Figure 1.** Selected moment maps of the Combined and Feathered Images for $^{12}$CO $J = 2-1$, $^{13}$CO $J = 2-1$, and C$^{18}$O $J = 2-1$. No streamer was detected in C$^{18}$O, and neither the disk nor the streamer was detected in SO. Left: peak intensity maps (moment 8). Middle: integrated intensity maps (moment 0). Right: line-of-sight velocity maps (moment 1). The color bars are centered at the line-of-sight velocity, 13.5 km s$^{-1}$, and peak intensity contours have been overlaid in white. The contours on the velocity maps depict peak intensity values ranging from 5 K to 20 K. The hatched ellipses in the lower left corners show the synthesized beam, and the horizontal line in the right corners show 1000 au for scale.

### 3.2. Mass of the Streamer

We estimate the mass of the streamer using the $^{13}$CO line emission and the nondetection of C$^{18}$O. We assume that the $^{13}$CO emission from the streamer is optically thin. To verify this, we estimate the optical depth of $^{13}$CO using the measured brightness temperature of $^{12}$CO, assuming it is optically thick. We find that $\tau(^{13}\mathrm{CO})$ is $\leqslant 0.1$ across the streamer. We estimate the number density of $^{13}$CO molecules in the $J = 2$ state with

$$n_{^{13}\mathrm{CO},\,J=2} = \frac{4\pi}{hcA_{21}} \int I\, dv \int d^2 d\Omega \tag{1}$$

where $h$ is the Planck constant, $c$ is the speed of light, $A_{21}$ is the spontaneous emission coefficient (F. L. Schöier et al. 2005),

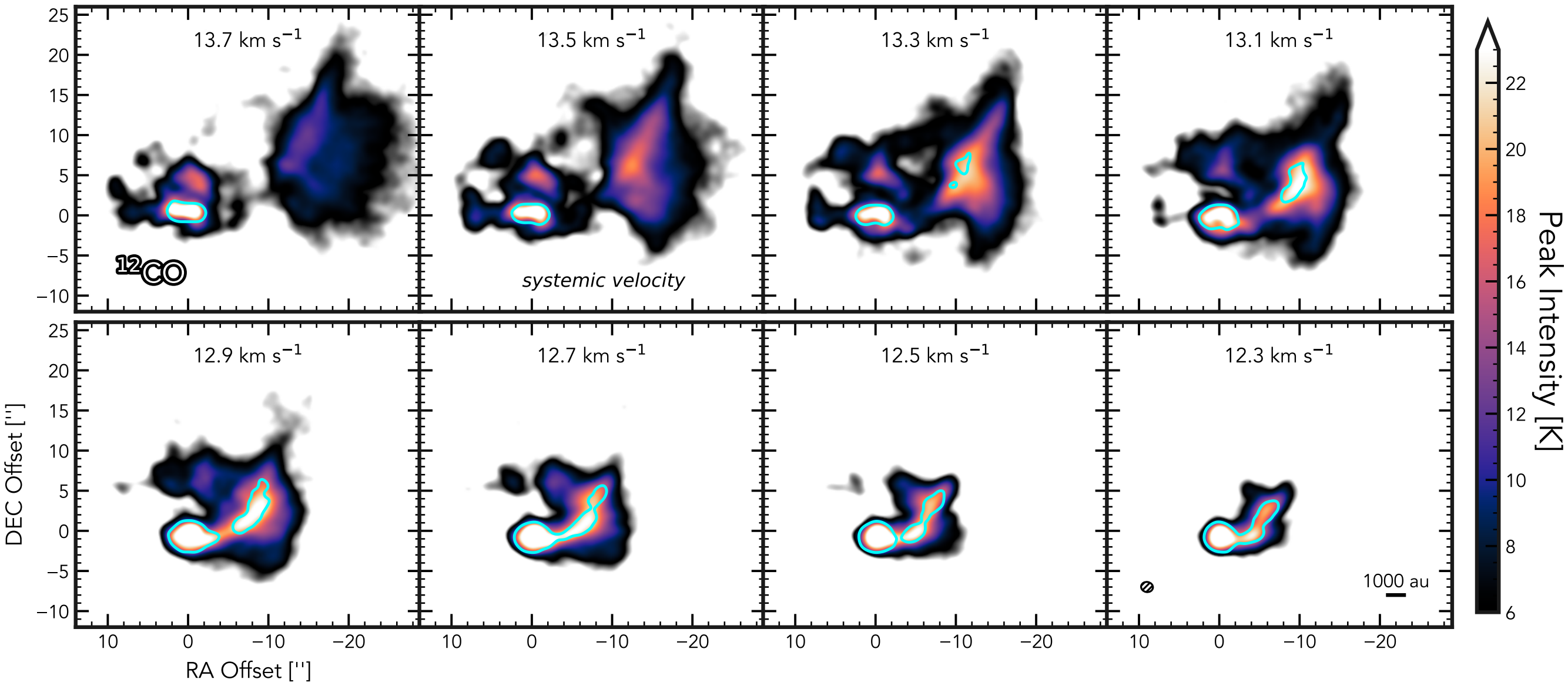

**Figure 2.** Selected channel maps for $^{12}$CO $J = 2-1$ emission. Emission from $^{13}$CO $J = 2-1$ at 4 K is overlaid in cyan. The synthesized beam is shown in the lower right panel, and the contour colors show the peak intensity in kelvins. The full channel maps for both molecular lines are shown in Appendix A.

$\int I\,dv$ is the integrated line intensity, $d$ is the distance to GW Ori, and $\Omega$ is the solid angle of the streamer. This allows us to derive the number density per pixel. We assume local thermodynamic equilibrium (LTE) for the $^{13}$CO emission, which is justified when the gas density exceeds the critical density of the transition ($n_{\rm crit} \sim 10^4$ cm$^{-3}$ for $^{13}$CO; J. G. Mangum & Y. L. Shirley 2015; M. L. R. van 't Hoff et al. 2018). For our simplistic estimate, this is a reasonable assumption, as other studies of streamers have found volume densities on the order of $10^4$–$10^5$ cm$^{-3}$ (J. E. Pineda et al. 2020), exceeding $n_{\rm crit}$, and LTE-based mass estimates are standard practice (for example, T. J. Thieme et al. 2022; M. T. Valdivia-Mena et al. 2022; A. Gupta et al. 2023). We calculate the line intensity using a moment-0 map of the Combined and Feathered data to which a Keplerian mask has been applied, which allows us to exclude the emission from the disk. To create the Keplerian mask, we use the `keplerian_mask` function from `GoFish` (R. Teague 2019). We assume a distance of 402 pc (Gaia Collaboration et al. 2018), an inclination $i = 37\overset{\circ}{.}9$ and a position angle (PA) $= -5°$ (measured eastward to the redshifted major axis from north; J. Bi et al. 2020; S. Kraus et al. 2020), and a total stellar mass of 5.3 $M_\odot$ (which includes the three central stars; I. Czekala et al. 2017; S. Kraus et al. 2020). We use a Keplerian mask radius of 3$''$.0 for $^{12}$CO and 2$''$.0 for $^{13}$CO. The Keplerian mask on the $^{12}$CO data is presented in Appendix A. We further mask the Keplerian-subtracted line intensity map by dropping any emission below 5$\sigma$, to isolate just the streamer emission. With this, we can estimate the total $^{13}$CO number density assuming local thermodynamic equilibrium (LTE):

$$n_{^{13}\rm CO,\,total} = n_{^{13}{\rm CO},J=2}\frac{Z}{2J+1}\exp\left[\frac{hB_eJ(J+1)}{kT}\right] \tag{2}$$

where

$$Z = \sum(2J+1)e^{-\frac{hB_eJ(J+1)}{kT}} \tag{3}$$

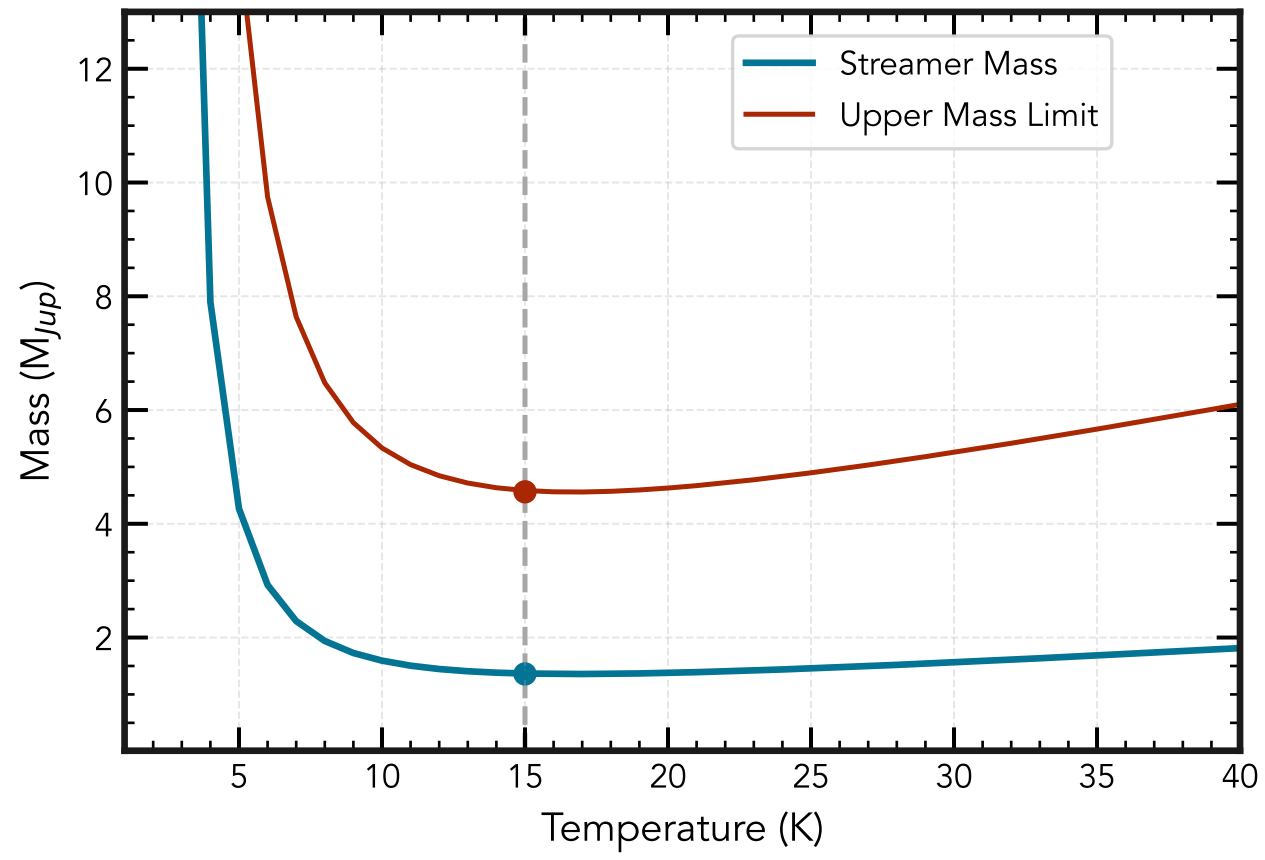

**Figure 3.** (Blue) Streamer mass estimate using the Keplerian-subtracted $^{13}$CO data and (red) upper mass limit using the nondetection of C$^{18}$O.

and $B_e$ = 55.10101 GHz is the rotational constant (H. S. P. Müller et al. 2005). We assume an interstellar CO abundance [$^{12}$CO]/[H$_2$] = $10^{-4}$ and a [$^{13}$CO]/[$^{12}$CO] abundance ratio of 1/60 (T. L. Wilson & R. Rood 1994). We note that the ratio of $^{12}$CO to H$_2$ can vary by an order of magnitude depending on the environment (D. Johnstone et al. 2003); most estimations agree that a magnitude of $10^{-4}$ is a reasonable assumption in molecular clouds (M. A. Frerking et al. 1982; J. L. Pineda et al. 2010; J. H. Lacy et al. 2017). To convert from number density to mass, we scale $n_{^{13}\rm CO,\,total}$ to $n_{\rm H_2}$, and then multiply this by $m_{\rm H_2}$.

We find the total mass of the streamer to be $1.6 \pm 0.8$ $M_{\rm Jup}$ ($\sim$0.001 $M_\odot$) assuming a gas temperature of 15 K, which is the median value of the $^{12}$CO $T_B$ values within the streamer. The assumed temperature does not significantly impact the mass estimate in the temperature regime of 10–40 K, as shown in Figure 3. Error analysis is conducted by taking the cube rms and adding/subtracting it to the 5$\sigma$ cutoff that is applied to the cube to isolate the streamer, such that the upper limit is defined as $5\sigma + {\rm rms_{cube}}$ and the lower limit is defined as $5\sigma - {\rm rms_{cube}}$.

The mass calculation is then repeated with these upper and lower limits to find the upper and lower errors. We also account for the 10% flux uncertainty discussed in Appendix B. The range of streamer masses when different gas temperatures are considered is shown in Figure 3. The 1.6 $M_{\rm Jup}$ streamer is similar to the combined mass of the outer two dust rings, which have masses of $\sim$168 $M_\oplus$ and $\sim$245 $M_\oplus$, or 1.3 $M_{\rm Jup}$ (J. Bi et al. 2020). The possibility that these dust rings are a second-generation disk created by the streamer is further discussed in Section 4.1.

We also compute an upper limit to the streamer mass using the nondetection of $C^{18}O$, following a similar process described above. We use Equation (2) to find the number density of $C^{18}O$, but approximate an upper limit by approximating $\int I\, dv$ as $5\sigma\ N\ \Delta v_c$, where $N$ is the number of channels with $^{13}CO$ streamer emission and $\Delta v_c$ is the spectral velocity channel width (J. G. Mangum & Y. L. Shirley 2015). We calculate $N$ by taking each channel where the $^{13}CO$ streamer emission is 5 times that of the $^{13}CO$ channel rms. This allows us to then find the number of $C^{18}O$ molecules. We take the ratio of $^{12}CO$ molecules to $C^{18}O$ molecules to be 560, and the ratio of $H_2$ molecules to $^{12}CO$ molecules to be $10^{-4}$ (T. L. Wilson & R. Rood 1994). We find an upper mass limit of $4.6^{+1.8}_{-0.98}$ $M_{\rm Jup}$ at 15 K, which is the $^{13}CO$ freeze-out temperature, shown in red in Figure 3. If we consider the observed $^{13}CO$ brightness temperatures, e.g., $<$10 K, the estimated mass increases slightly, but between 10 K and 40 K, it does not vary by more than a factor of 2.

Using the $^{13}CO$ and $C^{18}O$ number densities, we can estimate the column density. To derive this, we simply take the total number of molecules (for example, $n_{^{13}{\rm CO,\ total}}$) and divide it by the area of the streamer, which we derive by counting the number of pixels with emission. We find that $N(^{13}CO) = 2.8 \times 10^{16}$ cm$^{-2}$, $N(C^{18}O) = 9.8 \times 10^{15}$ cm$^{-2}$, and $N(H_2) = 1.65 \times 10^{22}$ cm$^{-2}$ at $T = 15$ K. These values are consistent with recent measurements in other streamer systems and their surrounding environments (C. Codella et al. 2024; M. Tanious et al. 2025; M. De Simone et al. 2026).

### 3.3. TIPSY Trajectory Fitting

To model the streamer, we utilize the open-source Python package TIPSY (Trajectory of Infalling Particles in Streamers around Young Stars; A. Gupta et al. 2024). TIPSY models the infall of material onto a central source in position–position–velocity (PPV) space, assuming the infall emission is a point cloud gravitationally interacting with the central protostar. TIPSY selects points along the streamer by dividing the emission into bins, which are determined by the user-input distance metric. For GW Ori, we assume a distance metric of 10, which is the default setting in TIPSY. From here, TIPSY computes intensity-weighted means and intensity-weighted standard deviations of the R.A., decl., and radial velocity (RV) values of all the points along the streamer. To fit the streamer with TIPSY, we make several edits to the Combined and Feathered Images. First, we apply a Keplerian mask with the same parameters discussed in Section 3.2. In addition to this, we clip the cube so that only the streamer is visible. These two steps ensure that we only consider infalling material in the fit, not the central Keplerian disk. We additionally limit the velocities to 10–15 km s$^{-1}$, discarding any emission outside of this range. The selected rms threshold is defined as `rms_thresh×median-absolute-deviation cube-flux` in TIPSY. This value can change the initial starting position of the point particle, and therefore introduces multiple solutions. To determine the best rms threshold, we perform a test by fitting the streamer with nine rms values ranging from 20 to 60. We find that the trajectories remain mostly consistent, with the biggest difference being the initial $x_0$, $y_0$ starting points. We opt to use a moderate rms of 40, which encompasses the main streamer component while maintaining a reasonable fitting fraction. We assume the same stellar parameters as in Section 3.2.

**Table 2**
Derived Properties from TIPSY Fit

| Parameter | $^{12}CO$ | $^{13}CO$ |
|---|---|---|
| $v_{xy0}$ [au] | 0.3 ± 0.1 | 0.3 ± 0.1 |
| $z_0$ [au] | 5000 ± 4292 | 4500 ± 4617 |
| Infall Time [Myr] | 0.048 ± 0.040 | 0.033 ± 0.033 |
| Mass Accretion Rate [$M_\odot$ yr$^{-1}$] | 3.6*E*−08 ± 0.000*e*+00 | 5.2*E*−08 ± 0.000*e*+00 |
| Specific Kinetic Energy [km$^2$ s$^{-2}$] | 0.05 ± 0.03 | 0.05 ± 0.04 |
| Specific Potential Energy [km$^2$ s$^{-2}$] | −0.6 ± 0.2 | −0.6 ± 0.2 |
| Specific Total Energy [km$^2$ s$^{-2}$] | −0.5 ± 0.2 | −0.5 ± 0.2 |
| Specific Angular Momentum [au km s$^{-1}$] | 1682 ± 1269 | 1703 ± 1345 |

The results of the TIPSY fit and additionally derived values for $^{12}CO$ and $^{13}CO$ are listed in Table 2. Figure 4 shows the best trajectory (dark red line) overlaid on the $^{12}CO$ and $^{13}CO$ peak intensity maps (left column). The orange trajectories show the next 100 best solutions, and the arrow shows the initial infall trajectory. The right column of Figure 4 shows the same TIPSY fits but in 3D space, demonstrating the projected $z$-distance of the trajectory. The positive $z_0$ values indicate that the streamer is further away from us than GW Ori, moving toward us. Figure 5 demonstrates the goodness of fit achieved by TIPSY for $^{12}CO$ and $^{13}CO$; the trajectory solutions for both molecules achieve a fitting fraction of one. Although there is only one best-fit solution, there are closely related trajectories, marked by the gray points. These alternative solutions are similar to the best-fit solution and do not lead to significantly different results.

The TIPSY fit predicts that the bulk of the streamer is behind the disk, at blueshifted velocities. From the observer's point of view, the streamer approaches the disk from behind, meeting it at the blueshifted region of the Keplerian disk. This scenario is consistent with the moment-1 maps presented in Figure 1. In this figure, the streamer components closest to the disk are blueshifted, but gradually become redshifted as we move further from the central star and toward the upper portion of the streamer.

TIPSY provides the initial LOS distance, the initial speed, and the initial direction on the plane of the sky (POS) for the infalling particle (A. Gupta et al. 2024). From these, we can derive the initial speed in R.A. and decl., as well as the initial LOS velocity offset. Following the methods outlined in A. Gupta et al. (2024), we derive the infall timescale, infall rate, specific kinetic energy, specific gravitational potential energy, total energy (TE), and angular momentum. Table 2 summarizes these quantities, and we discuss these results in Section 4.

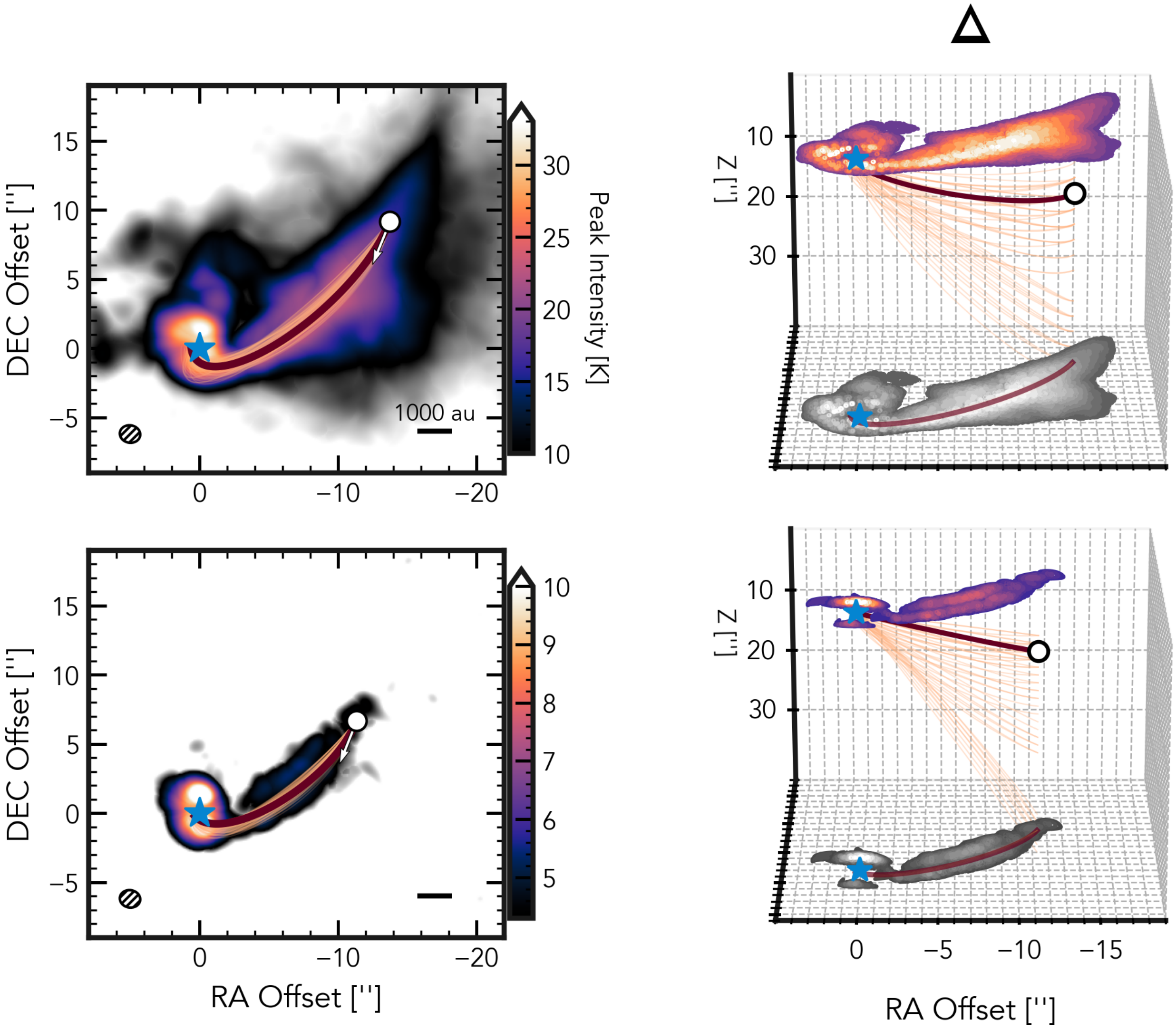

**Figure 4.** Left column: TIPSY fit plotted in dark red over peak intensity maps of $^{12}$CO and $^{13}$CO. The light orange lines show the next best 100 solutions. The white arrows show the initial velocity of the point particle. The beam size and 1000 au scale are shown in the left and right corners, respectively. Right column: the same TIPSY fits visualized in 3D space with a colored peak intensity map projected onto the $x$–$y$ plane at $z = 0$, and a binary colormap of the object projected at $z = 40$. The triangle represents the location of the observer, which is at $z \ll 0$. The complete set of TIPSY fitting results for $^{12}$CO and $^{13}$CO is available as data behind the figure.

(The data used to create this figure are available in the online article.)

### 3.4. Position–Velocity Space

Figure 6 shows position–velocity diagrams for $^{12}$CO and $^{13}$CO emission. As opposed to taking a thin $p$–$v$ slice across the disk, we opt to collapse the entirety of the emission across the disk axes so that we encompass the total extent of the streamer. The disk surface in the upper left corner of plot (A) demonstrates the collapse across the disk major axis (plots (A) and (B)). Plot (C) shows the position–velocity diagram along the best-fit trajectory, which is shown in red. In plots (A) and (B), we overplot the bounds of Keplerian rotation in white, given by $v_{\rm Kep} = \sqrt{GM_*/r} \times \sin(i)$, using the source properties discussed in Section 3. The Keplerian rotation of the central disk is visible, with the large non-Keplerian component of the streamer visible in the bottom right of the plots.

In panels (A) and (B), we additionally plot the escape velocity, or gravitational boundary, in yellow, given by $v_{\rm esc} = 2 \times v_{\rm Kep}$. The streamer component emission is within the gravitational boundaries, consistent with the TE predictions from TIPSY. If TE $< 0$, the gas is in a bound elliptical orbit; TE $= 0$ implies a bound parabolic orbit; and TE $> 0$ implies an unbound hyperbolic orbit. TIPSY finds that both molecular tracers have TE $< 0$, suggesting that the gas within the streamer is in a bound elliptical orbit, consistent with what we see in the $p$–$v$ diagram. In panel (C), the coherent flow of the streamer can be seen in both $^{12}$CO and $^{13}$CO; the emission traces the best-fit trajectory shown in red. The streamer is consistently blueshifted as it meets the southern end of the disk, with velocities ranging from $\sim$13.5 km s$^{-1}$ (the systemic velocity) down to $\sim$10 km s$^{-1}$. The smooth velocity gradients in the $p$–$v$ diagram imply that, in addition to apparent spatial coherence, the streamer is a kinematically coherent structure.

Two distinct components of the streamer are visible in panels (A), (B), and (D) of Figure 6. The first, which we will designate as the "high-velocity" component (as the velocity is greater than the systemic velocity), begins just above the systemic velocity of 13.5 km s$^{-1}$ in $^{12}$CO in panel (A). Following the stream downward, it breaks into a lower section (the "low-velocity" component) at ∼10″ in both $^{12}$CO and $^{13}$CO. The components are labeled in panel (A), in position–velocity space, and in panel (D), in position–position space. The two streams meet the disk at slightly different radial positions, with the high-velocity stream meeting the disk closer in than the low-velocity component, which appears to connect with only the outer region of the disk. The origin of the structure is unknown, and we discuss the possibilities in Section 4.2.

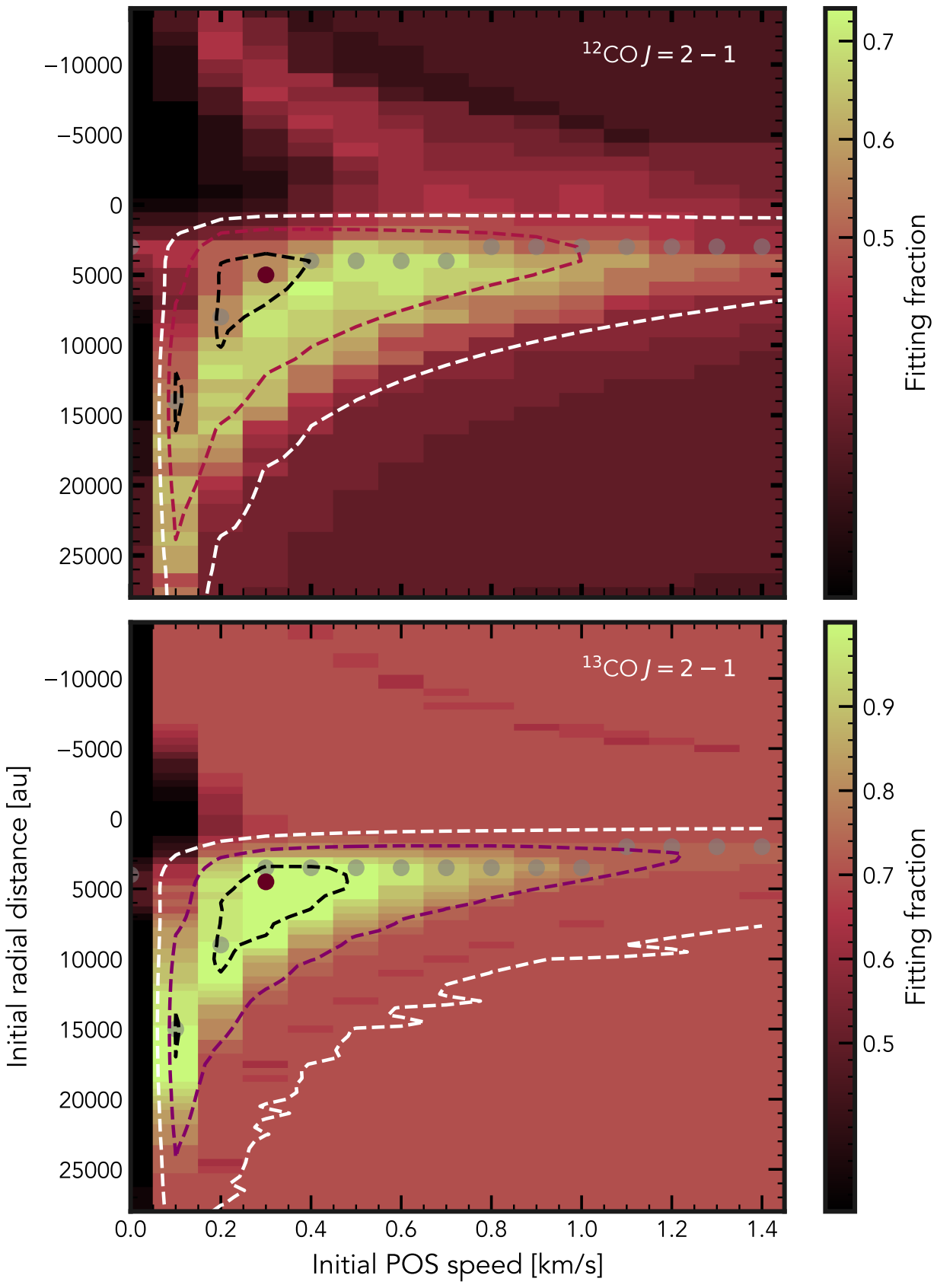

**Figure 5.** Fitting fraction (goodness of fit) as a function of the initial LOS distance and the initial plane-of-the-sky (POS) speed. The red dot shows the best-fit solution, while the gray dots show the best-fit solution in each POS speed bin. The contours show the 2nd, 10th, and 25th percentiles of the standard deviation of the filling fraction. Positive initial radial distances are further from the observer, while negative ones are closer.

### 3.5. Infall Timescales and Mass Infall Rate

The TIPSY model estimates an infall time for the material, defined as the time taken for the best-fit solutions to reach the point closest to the protostar, starting from the farthest point in the observed streamer; this value is dependent on the initial starting point and the derived velocities. For $^{12}$CO and $^{13}$CO, TIPSY finds infall times of 0.048 ± 0.040 Myr and 0.033 ± 0.033 Myr, respectively; these values are much lower than the estimated system age of 0.3–1.3 Myr (I. Czekala et al. 2017). We can compare this to the freefall time, given by $t_{\rm ff} \sim \sqrt{d^3/GM_{\rm tot}}$ where $d$ is the radial distance from a point particle (at the largest extent of the streamer) to the central star and $M_{\rm tot}$ is the total mass of the system. The freefall time represents a lower bound on the infall time. Total mass is given by stellar mass (5.3 $M_\odot$) plus disk mass (0.12 $M_\odot$). If we only consider the on-sky radial distances, $x$ and $y$, we find a freefall time of 0.03 Myr for both molecular tracers, comparable to what TIPSY finds. These infall timescales are broadly consistent with those of other streamers, such as Per-emb-50 (∼0.02 Myr; M. T. Valdivia-Mena et al. 2022), MGM 2012 (0.05 Myr; L. Cacciapuoti et al. 2024), and Per-emb-2 (0.01 Myr; J. E. Pineda et al. 2020), as well as those found in numerical simulations, ∼0.01 Myr (J. Mauxion et al. 2024; J. Calcino et al. 2025; L.-A. Hühn & C. P. Dullemond 2025). Other streamers have much shorter estimated infall times, such as S CrA and HL Tau (0.008 Myr, 0.003 Myr; A. Gupta et al. 2024), but we note that measurements of this kind are also dependent on the observational sensitivity.

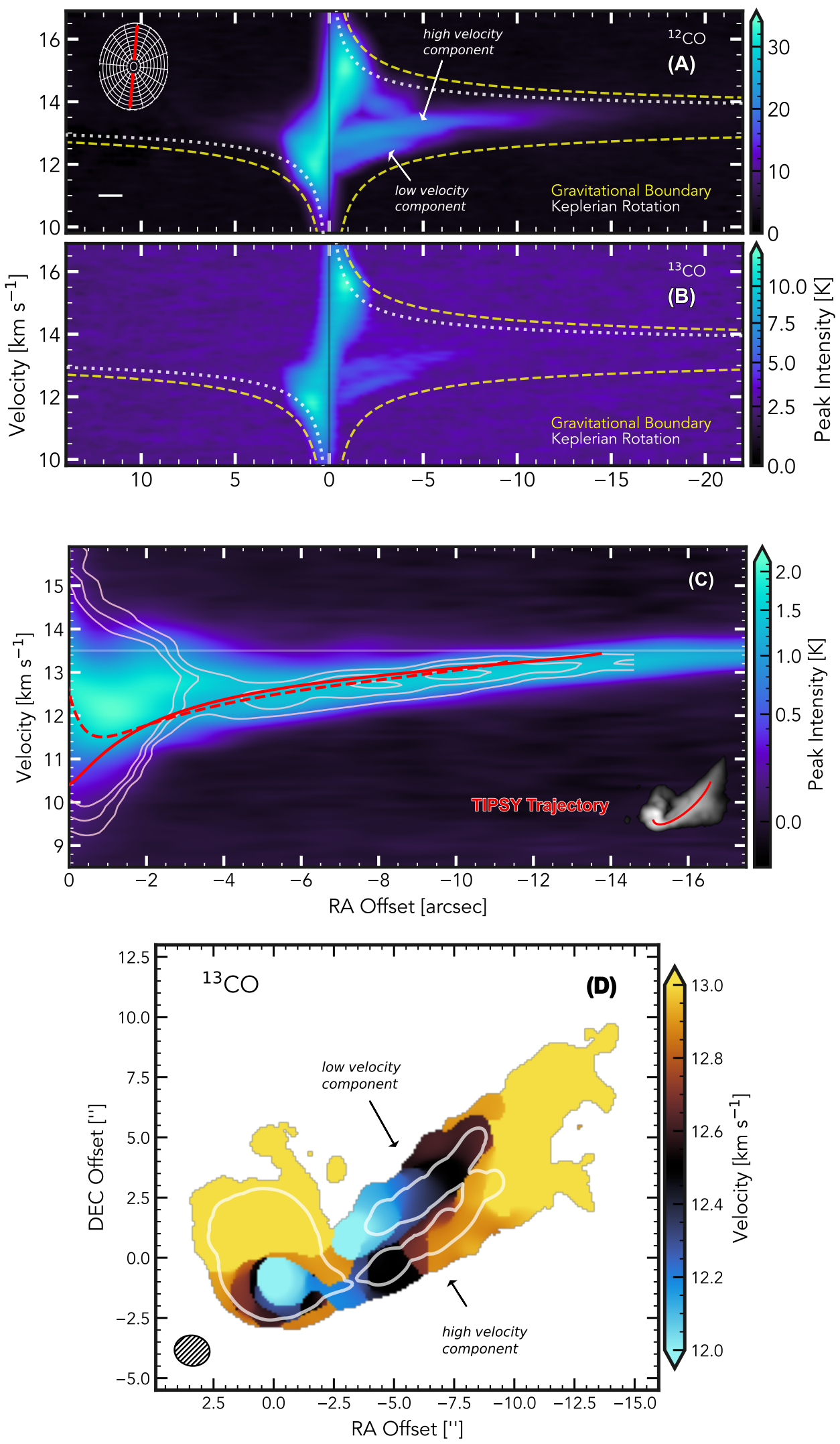

**Figure 6.** (A)–(B) Position–velocity diagram collapsed across the disk major axis of the data cube for $^{12}$CO (top) and $^{13}$CO (bottom). The white dotted lines show the expected bounds of Keplerian rotation, and the yellow dashed lines show the gravitational boundary, where any emission outside of those lines is considered gravitationally unbound. The two velocity components are annotated with arrows. The beam size is shown as a white line in the bottom left corner of (A). (C) Position–velocity diagram collapsed across the TIPSY trajectory (shown in red) for $^{12}$CO and $^{13}$CO ($^{13}$CO in pink contours). (D) Velocity map of $^{13}$CO disk and streamer, zoomed in and color-stretched to highlight the high- and low-velocity components.

We can estimate the mass infall rate as $\dot{M}_{\rm in} = M_{\rm streamer}/t_{\rm infall}$, where $M_{\rm streamer}$ is the streamer mass calculated in Section 3.2, equal to 1.6 $M_{\rm Jup}$. We use the infall times provided by TIPSY, given in Table 2. Using this prescription, we find a mass infall rate of $(3.64 \pm 3.19) \times 10^{-8}$ $M_\odot$ yr$^{-1}$; the large errors are driven by the uncertainty in the infall timescales. Our derived mass infall rate is lower than many derived for other streamers, for example, M512 with an infall rate of $1.5 \times 10^{-6}$ $M_\odot$ yr$^{-1}$ (L. Cacciapuoti et al. 2024), L1489 IRS with $5 \times 10^{-7}$ $M_\odot$ yr$^{-1}$ (H.-W. Yen et al. 2014), Lupus 3-MMS with $\sim 1 \times 10^{-6}$ $M_\odot$ yr$^{-1}$ (T. J. Thieme et al. 2022), and Per-emb-50 with $1.3 \times 10^{-6}$ $M_\odot$ yr$^{-1}$ (M. T. Valdivia-Mena et al. 2022). Additionally, GW Ori's streamer infall rate is about an order of magnitude lower than the stellar accretion rate estimated from optical and UV data of $\dot{M}_{\rm acc} \approx 3 \times 10^{-7}$ $M_\odot$ yr$^{-1}$ (N. Calvet et al. 2004; M. Fang et al. 2014). Because the material fed by a streamer typically lands on the outer disk and must be accreted to the inner edge of the disk before it reaches the star, the larger stellar accretion rate suggests that the streamer infall rate could have been much larger in the past and that we are now witnessing the end stage of the streamer, although the stellar multiplicity may also play a role here.

### 3.6. Angular Momentum

A central question about late-stage infall is whether the accreting streamer can perturb a disk enough to tilt or misalign it (see I. Thies et al. 2011; M. Kuffmeier et al. 2021). To explore this scenario in GW Ori, we estimate the angular momentum delivered by the streamer. To derive the specific angular momentum of the streamer, we take the cross product of $\boldsymbol{r_0}$, the position vector, and $\boldsymbol{v_0}$, the velocity vector, both derived from the TIPSY fit. We find a specific angular momentum of $1682 \pm 1269$ au km s$^{-1}$ for $^{12}$CO and $1703 \pm 1345$ au km s$^{-1}$ for $^{13}$CO (see Table 2). The large uncertainties are mostly due to the large uncertainty on $z_0$, the initial starting point along the LOS (see right column of Figure 4, which shows the multiple $z_0$ solutions in orange). Despite these large uncertainties, the value is almost always lower than that of the disk. We compare these specific angular momentum values to the disk momentum, which is computed using $l(r) = \sqrt{GM_* r}$ where $M_*$ is the stellar mass and $r$ is the disk radius. At the disk edge ($r = 3''$ or $\sim$ 1206 au), we find a specific angular momentum of 2400 au km s$^{-1}$, so $j_{\rm disk} > j_{\rm streamer}$. This suggests that once the streamer meets the disk, the infall material will be likely transported inward. Simulations from V.-M. Pelkonen et al. (2025) show a positive correlation between the angular momentum of streamers formed by Bondi–Hoyle (BH) accretion, $j_{\rm BH}$, and stellar mass: $j_{\rm BH} = 4 \times 10^{20}$ cm$^2$ s$^{-1}$ $(M/1\ M_\odot)^{1.1}$. Our derived streamer angular momentum is consistent with this relation; for a 5.3 $M_\odot$ system, $j_{\rm BH} \approx$ 1600 au km s$^{-1}$, while we find a streamer angular momentum of 1682 au km s$^{-1}$. In Section 4.3, we will discuss BH accretion in more detail. Figure 7 summarizes these findings, showing the analytically derived disk angular momentum (black line) with the estimated streamer angular momentum and associated errors in red, and the angular momentum of streamers formed by BH accretion in purple.

We can also calculate the total angular momentum of the streamer and compare it to that of the disk. The streamer angular momentum is estimated by multiplying the specific angular momentum by the streamer mass of 1.6 $M_{\rm Jup}$, which results in $L_{\rm streamer} = {\sim}2500 \pm 1800$ $M_\odot$ au km s$^{-1}$. The disk angular momentum is given by $L_{\rm disk} = \int_0^{R_{\rm max}} l(r) dM(r)$, where $dM = 2\pi\Sigma(r)dr$ and $\Sigma(r) = \Sigma_0 \left(\frac{r}{r_0}\right)^{-1}$. We assume a disk mass of 0.12 $M_\odot$ (M. Fang et al. 2017) and an outer radius of $R_{\rm max} = 1200$ au, which results in $\Sigma_0 \simeq 1.43$ g cm$^{-2}$ at $r_0 = 100$ au. Using this prescription, we find that the disk total angular momentum is $\sim 1.9 \times 10^4 M_\odot$ au km s$^{-1}$. This suggests that at the present time, the streamer does not have sufficient angular momentum to significantly perturb the alignment of the disk.

However, as we have mentioned previously, we cannot rule out the possibility that we are observing the end stages of the infall event, where the angular momentum added to the disk by the streamer in the past could have caused the disk misalignment. For this to be the case, the streamer must have delivered at least 180 $M_{\rm Jup}$, or 0.17 $M_\odot$, to the disk over its lifetime assuming that the specific angular momentum has been constant. This estimate is larger than the disk mass (0.12 $M_\odot$), implying that a significant amount of the disk mass was delivered via the streamer. MHD simulations predict that some stars may gain upward of 50% of their mass from late infall (M. Kuffmeier et al. 2023), while numerical simulations suggest that $\sim$13% of initial cloudlet material will end up in the disk in an accretion scenario (L.-A. Hühn et al. 2025). If we assume that the current streamer has just 15% of the initial cloudlet material, we estimate that the original cloudlet mass would have been about 2 $M_\odot$. This is consistent with observational estimates of dense cores and filaments in Orion (D. Polychroni et al. 2013), as well as streamer masses estimated for younger Class 0/I sources (J. E. Pineda et al. 2020).

If the streamer angular momentum does not match that of the disk edge, the material accreting to the disk must move inward until it reaches a stable orbit. We can estimate this radius, denoted as the centrifugal or circularization radius $R_c$, as $\frac{j_c^2}{GM_*}$, where $j$ is the specific angular momentum of the infalling material (R. K. Ulrich 1976). We find a circularization radius of $\sim$500 au, meaning that the material accreted from the streamer will move inward in the disk toward the central star until this radial location is reached. The mismatch between the disk outer radius and the centrifugal radius means there must be a transition zone between the two, which can aid in the development of dynamical perturbations, such as spirals or vortices (G. Lesur et al. 2015; A. Kuznetsova et al. 2022). Instabilities like RWI will produce vortices and pressure bumps, which can drive disk substructures such as spirals, rings, and gaps. Although the outermost continuum ring of GW Ori is inward of the centrifugal radius, evidence of RWI or other instabilities may be found via gas kinematics, and giant planets can still potentially form outside of the continuum regions. Future work using high angular resolution observations of the central disk could help shed light on disk regions undergoing angular momentum transport.

## 4. Discussion

### 4.1. The Streamer as the Driver of Disk Misalignment

Simulations suggest that infall could be responsible for disk misalignment, which can produce shadows seen in a growing number of scattered light observations (M. Benisty et al. 2017, 2018; S. Casassus et al. 2018; C. Ginski et al. 2021), including in GW Ori (S. Kraus et al. 2020). There are several plausible explanations for the disk misalignment in GW Ori, including gravitational torque of the triple stellar system (D. Lai & F. Foucart 2014; J. Bi et al. 2020; S. Kraus et al. 2020; A. K. Young et al. 2023), gravitational interaction between a forming planet and the disk (K. Batygin 2012; T. Matsakos & A. Königl 2017; J. L. Smallwood et al. 2021), and primordial misalignment imparted by differences in the angular momentum of the prestellar core and the parental cloud (M. Kuffmeier et al. 2024; V.-M. Pelkonen et al. 2025). Here, we explore the latter scenario using the derived estimates of the disk and streamer dynamical properties.

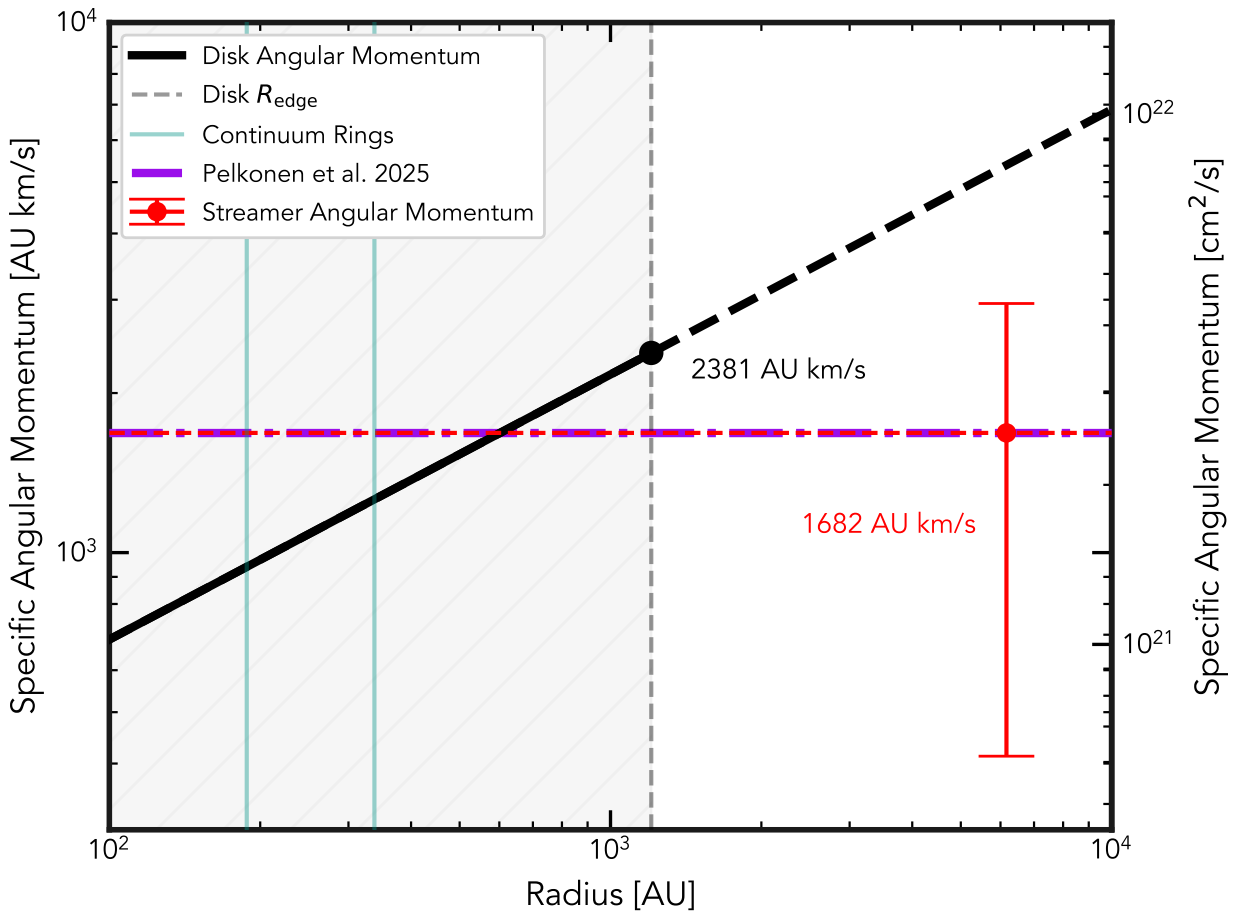

**Figure 7.** Specific angular momentum ($j$) versus radius for the GW Ori disk and the streamer. The black line is the disk $j$ estimated using $l = \sqrt{GM_{*}R_d}$; it is dashed past the gaseous disk edge. The streamer angular momentum derived from TIPSY is shown in red, with the error region shaded. The purple line shows the theoretical specific angular momentum of streamers formed via BH accretion, measured using Equation (10) from V.-M. Pelkonen et al. (2025). The solid lines inward of 1000 au are the radial locations of the outermost continuum rings.

Although we find the streamer's angular momentum is smaller than that of the disk presently (Section 3.6), examining the angular momentum vectors of the disk and infall components can reveal whether or not the streamer could be responsible for the misaligned structure of GW Ori's triple dust disk. We obtain the angular momentum vectors of the continuum rings in 3D space using $\boldsymbol{v} = r_{\rm vec}\,\hat{v}$, equivalent to $r_{\rm vec}\,(\sin i \cos {\rm PA}\,\hat{x} + \sin i \sin {\rm PA}\,\hat{y} + \cos i\,\hat{z})$. We assume the geometric parameters from J. Bi et al. (2020), who find the inner disk has an inclination of $\sim 11^\circ$, the middle disk has an inclination of $\sim 35^\circ$, and the outer disk has an inclination of $\sim 40^\circ$. The streamer angular momentum vector is calculated using the $\boldsymbol{r}$ and $\boldsymbol{v}$ vectors from the TIPSY fit. Figure 8 shows the vectors of the streamer (maroon) and the three dust rings (orange, green, and blue). The angular momentum vectors point toward positive $z$-values, i.e., away from the observer, as the disk has clockwise rotation (I. Czekala et al. 2017; J. Bi et al. 2020). Here, the streamer trajectory and disk scales are for illustrative purposes, and do not represent the true size scale of the system. We have also rotated the angular momentum vectors toward the observer for a clearer viewing angle. We find that the angular momentum vector of the streamer is most closely aligned with that of the outer dust ring, with a relative inclination angle of only 3°, compared to 8° for the middle dust ring and 32° for the inner dust ring.

The alignment between the streamer and the outer ring within 3° provides strong observational evidence that the streamer is the source of the disk misalignment. Probabilistically, the likelihood of this alignment happening randomly can be estimated as $P(\theta) = \frac{1}{2}\int_0^\theta \sin\theta d\theta$; this gives a probability of ≈0.07%, meaning that the random chance of alignment is slim. These results support a scenario in which the streamer contributes to the reorientation of the disk, as predicted by hydrodynamic simulations where asymmetric accretion from infalling material can disrupt the original orientation of the disk (M. Kuffmeier et al. 2021, 2023; V.-M. Pelkonen et al. 2025). In these simulations, the infall mass can be as little as $10^{-4}\ M_\odot \simeq 0.1\ M_{\rm Jup}$ to cause this disruption. If the angular momentum of the infall is sufficiently large, the material may not be efficiently transported, allowing the disk to grow and create a potentially misaligned second-generation disk. This scenario has even been observed in the younger Class I protostar L1489 IRS (M. Tanious et al. 2024, 2025), where a streamer has contributed to a warped disk. In contrast, the significant misalignment between the outer (and middle) dust ring and the innermost ring further supports the idea that the outer disk could be a second-generation disk, assembled from the infalling material at a later stage than the central disk component. Such an evolutionary pathway is consistent with simulations showing that disks can undergo episodic rebuilding, where streamers form an outer disk with a distinct orientation from the preexisting inner disk (I. Thies et al. 2011; C. P. Dullemond et al. 2019; M. Kuffmeier et al. 2020); see also the observational work of M. Tanious et al. (2024, 2025) and X. Mai & X. Liu 2025). The combination of streamer–disk alignment, inner–outer disk misalignment, and ongoing late-stage infall in GW Ori therefore provides a compelling observational example of streamer-driven disk misalignment and rebuilding in a Class II system.

As the streamer accretes onto the disk, we expect an impact zone where the bulk of the infalling material meets the disk. Using the TIPSY models, we can estimate the radial location of this impact zone by finding the $x$ and $y$ coordinates where $z = 0$. For the best-fit TIPSY solution, we find the impact zone at $x \simeq -0\rlap{.}''082$ and $y \simeq -0\rlap{.}''87$. The results are shown in Figure 9, overlaid on a continuum image of GW Ori, with the next best five solutions shown in orange. All of the potential infall zones lie around the outermost or middle dust ring, with the best infall solution closely aligned with the outermost dust ring.

As discussed previously, simulations suggest that late-stage infall can produce pressure bumps or even seed second-generation disks (C. P. Dullemond et al. 2019; M. Kuffmeier et al. 2020). If the infall zone truly lies within the radial ranges shown in Figure 9, it is reasonable to speculate that the accreting material created a pressure maximum at that location, aiding in the creation of the outer dust ring. The masses of the dust rings are measured to be ∼7, ∼168, and $\sim 245\ M_\oplus$, respectively (J. Bi et al. 2020). If we assume the middle and outer rings are comprised entirely of fresh material from the streamer, we can reestimate the infall rate by taking the mass of the dust rings divided by the infall time determined in Section 3.5. Taking $M_{\rm gas} \approx 100\ M_{\rm dust}$ and dividing by $3.64 \times 10^{-8}\ M_\odot\ {\rm yr}^{-1}$, we estimate that the infall has been taking place for 3.4 Myr, assuming the infall rate has stayed the same. This further supports the idea that we are witnessing the end stages of the infall.

This theory is corroborated by the fact that the outer dust ring and streamer angular momentum vectors are closely aligned. Isotopic evidence from outer disk material in our solar system shows a clear chemical distinction between inner and outer disk reservoirs, indicating that these regions experienced different chemical and dynamical histories (E. van Kooten et al. 2024; E. van Kooten & S. Lodal 2026). A similar mechanism may be operating in GW Ori, where streamer-fed infall is introducing chemically distinct material into the outer disk.

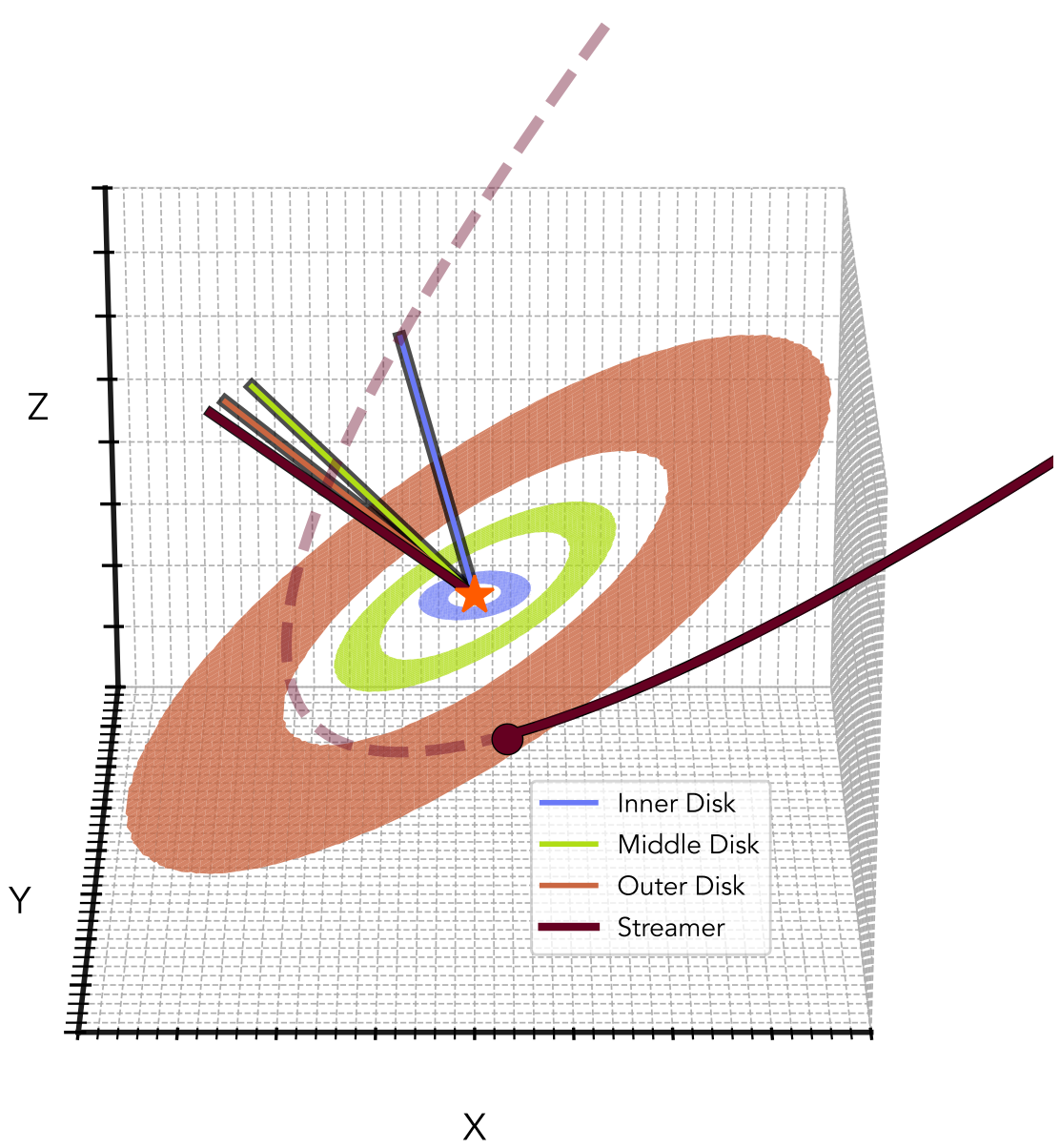

**Figure 8.** Visualization of GW Ori's three dust rings, with inclination and PAs from J. Bi et al. (2020), and the streamer trajectory (maroon) in 3D space with the angular momentum vectors of each. Not to scale; the sizes of the rings are relative. The angular momentum vectors have been inverted for visual clarity. The streamer angular momentum is closely aligned with that of the outer disk, suggesting that the infall is the source of the disk misalignment. The dashed line shows the remainder of the trajectory once it has met the disk.

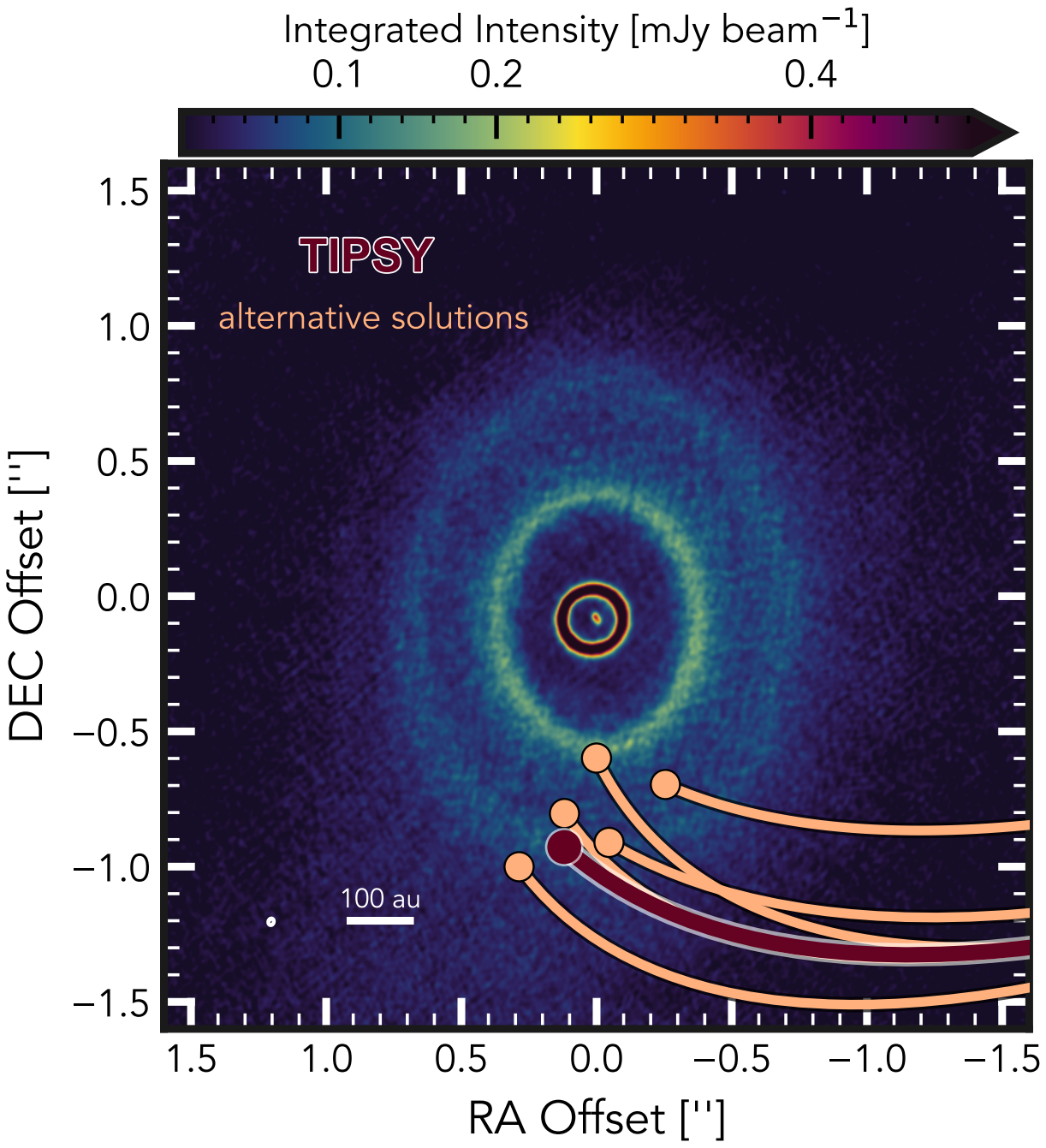

**Figure 9.** TIPSY trajectories and their points of impact overlaid on 1.3 mm dust continuum image of GW Ori from S. Kraus et al. (2020). The best-fit TIPSY solution is shown in maroon, with the next best five solutions shown in orange. The continuum beam size and a 100 au scale are shown in white in the lower left corner.

### 4.2. Infall Structure

The streamer has a prominent intensity drop through its center, visible in both $^{12}$CO and $^{13}$CO. This structure is especially distinct in the peak intensity maps, centroid velocity maps, and position–velocity diagrams (see Figures 1 and 6). Here, we explore several possibilities for the origin of this structure.

One possibility is spatial filtering. In AB Aur, J. Speedie et al. (2025) observed a bifurcation, or fork, in $^{12}$CO emission, centered at the systemic velocity. However, in our case, the split in emission is not centered about a velocity, and persists from $\sim$11 km s$^{-1}$ up to $\sim$13 km s$^{-1}$. Additionally, our observations are spatially well sampled, as we have combined pointings from the 12 m, 7 m, and Total Power arrays that probe different spatial scales. Therefore, we do not attribute this effect to spatial filtering.

The split structure could instead be explained if the streamer can be approximated by an optically thin hollow cylinder. The streamer has a low brightness temperature, generally $\lesssim$20 K, which is below the freeze-out temperature of CO. It is therefore reasonable to assume that most of the $^{13}$CO emission we are seeing is optically thin, as discussed in Section 3.2. If this is the case, an optically thin cylinder should exhibit lower brightness temperature in its center as compared to the edges because the LOS through the outer rim will intersect a longer path length of material than the LOS through the center. Other possibilities are that the edges of the cylinder experience increased shear, leading to shock or compressional heating, or that the cylinder simply has a low-density interior, or that the interior of the streamer is cold.

The velocity structure of the streamer can further explain the observations if we consider a rotating cylinder. Figure 6 demonstrates the "high-velocity" and "low-velocity" components of the infall. This structure can also be seen in the LOS velocity maps, particularly in $^{13}$CO (see panel (D) of Figure 6). If the streamer can be approximated as a rotating cylinder, we would expect two different velocity components as one side rotates toward the observer and the other away from the observer. This may manifest as multiple different velocity components, just as we see. Panel (D) of Figure 6 shows the centroid velocity map of $^{13}$CO emission, highlighting that the edges of the streamer have slightly different velocities. As the velocity difference appears to be $\sim$500 m s$^{-1}$, this would imply a velocity of around 250 m s$^{-1}$ for either side of the cylinder, which is quite high relative to the expected sound speed at temperatures of 20 K (e.g., $\sim$270 m s$^{-1}$). Numerical simulations of infall, together with future observations of an additional $^{13}$CO transition to better constrain the streamer's temperature and column density, will allow us to explore this rotating cylinder scenario.

Finally, infall or cloudlet accretion processes may naturally form multiple streamers. Class 0/I sources such as IRAS 04239 +2436 (J.-E. Lee et al. 2023), Lupus 3-MMS (T. J. Thieme et al. 2022), and Oph IRS 63 (C. Flores et al. 2023) have two or even three streamers. Alongside these observations, simulations have predicted similar infall structure. Hydrodynamic simulations by L.-A. Hühn & C. P. Dullemond (2025) saw multiple streamers arise over the course of the infall lifetime, which then later merge; Y. Yano et al. (2024) revealed similar multistream structure in core-collision simulations. A double-tail structure is also seen in hydrodynamic simulations from C. P. Dullemond et al. (2019). In fact, additional emission that is not associated with the disk or main streamer can be seen in the $^{12}$CO peak intensity map to the north, or northwest, of the disk. This additional material has a distinct velocity, different from that of the streamer by $\sim$1 km s$^{-1}$. This emission is visible in the $^{12}$CO $p$–$v$ diagram (Figure 6, panel (A)), at $\sim$14 km s$^{-1}$, above the high-velocity component. It is thus possible that this additional emission is part of another streamer, or the streamer we see now is/was multiple filaments that have merged over time as the accretion process comes to an end.

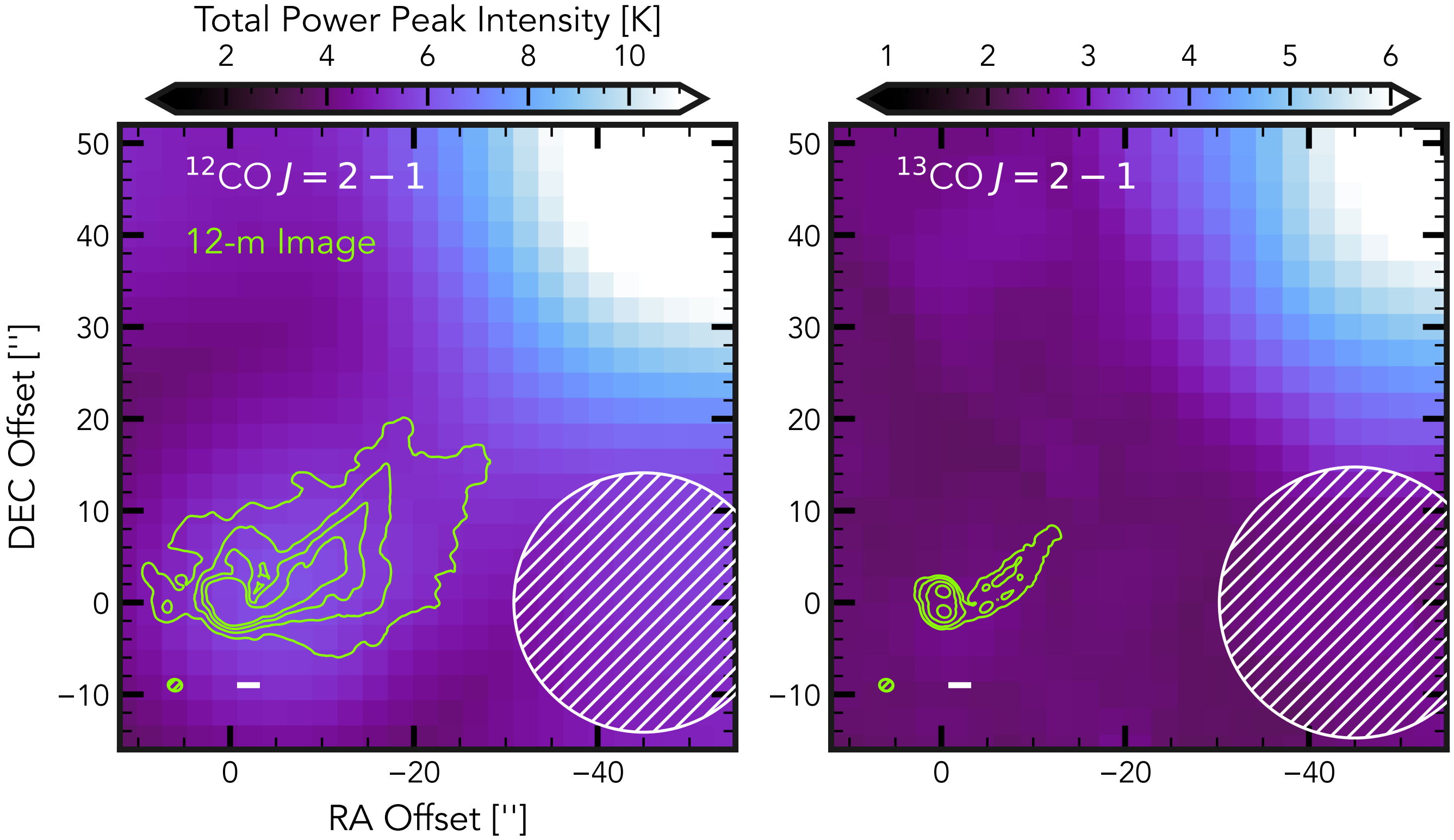

**Figure 10.** Peak intensity maps of the Total Power images with contours of the disk and streamer from the Combined and Feathered Images in green. The beam sizes are shown in the left and right corners of the plot with the respective colors. A white bar demonstrating 1000 au is shown at the bottom of the plot.

### 4.3. The Origin of the Streamer: Bondi–Hoyle Accretion?

Having established that the streamer's angular momentum vector is tightly aligned with the outermost dust ring and that the current infall rate is low, we now turn to the questions of where the streamer came from and how the system may evolve in the future. We propose that the streamer is fed by BH accretion from the surrounding molecular cloud, evidenced by the Total Power data, and that the infall event is approaching its end.

#### 4.3.1. Connection to the Surrounding Cloud

GW Ori is located within the massive star-forming region of Orion. Here, we explore the streamer in this larger context. The Total Power observations reveal large-scale emission toward the upper right of the streamer (northwest), not previously detected in the archival data. This emission is highlighted in Figure 10, visible in the top right corner of the $^{12}$CO and $^{13}$CO FOVs (at approximately $-50''$, $50''$). The disk and streamer are shown as green contours for scale. The bright emission is $\sim$70$''$ on-sky from the central disk, and $\sim$20$''$ on-sky from the furthest extent of the streamer seen in the $^{12}$CO Combined Image. Figure 11 shows three spectra, taken from the $^{12}$CO Total Power image. The spectra are taken at three locations: toward the disk (Region 1, orange), toward the streamer (Region 2, blue), and toward the cloud (Region 3, black); the locations that the spectra are drawn from are shown as circles in the inset plot. It is clear that the disk and streamer are closely related in velocity space, with a $\Delta v_{\rm los}$ of 0.1 km s$^{-1}$. The streamer and the cloud are separated by another $\Delta v_{\rm los}$ of 0.4 km s$^{-1}$. The emission at $\sim$9 km s$^{-1}$ is likely contamination and not associated with the streamer. Although we cannot know the true 3D distance between the streamer and the cloud, the velocities and the TIPSY solutions imply that there is not a large physical separation. It is therefore plausible that the infall we see in the high-resolution data is still physically connected to the larger mass reservoir of the local star-forming region.

Zooming out further, Figure 12 shows a 100 $\mu$m observation of a patch of the Orion star-forming region (IRAS; M.-A. Miville-Deschênes & G. Lagache 2005). At these scales, the environment is not isolated; it is crowded with interstellar material, filamentary structures, and dense cores. In this setting, many of our findings can be reasonably approximated within a BH accretion framework, in which a gravitationally bound system accretes material from the surrounding medium as it moves through a structured environment (H. Bondi & F. Hoyle 1944). This process can naturally give rise to late-stage infall, providing a viable mechanism for delivering mass and angular momentum to the disk at late times (P. Padoan et al. 2005, 2025; A. J. Winter et al. 2024).

The BH radius, $R_{\rm BH}$, is written as

$$R_{\rm BH} = \frac{2GM_*}{c_s^2 + v_{\rm rel}^2}, \tag{4}$$

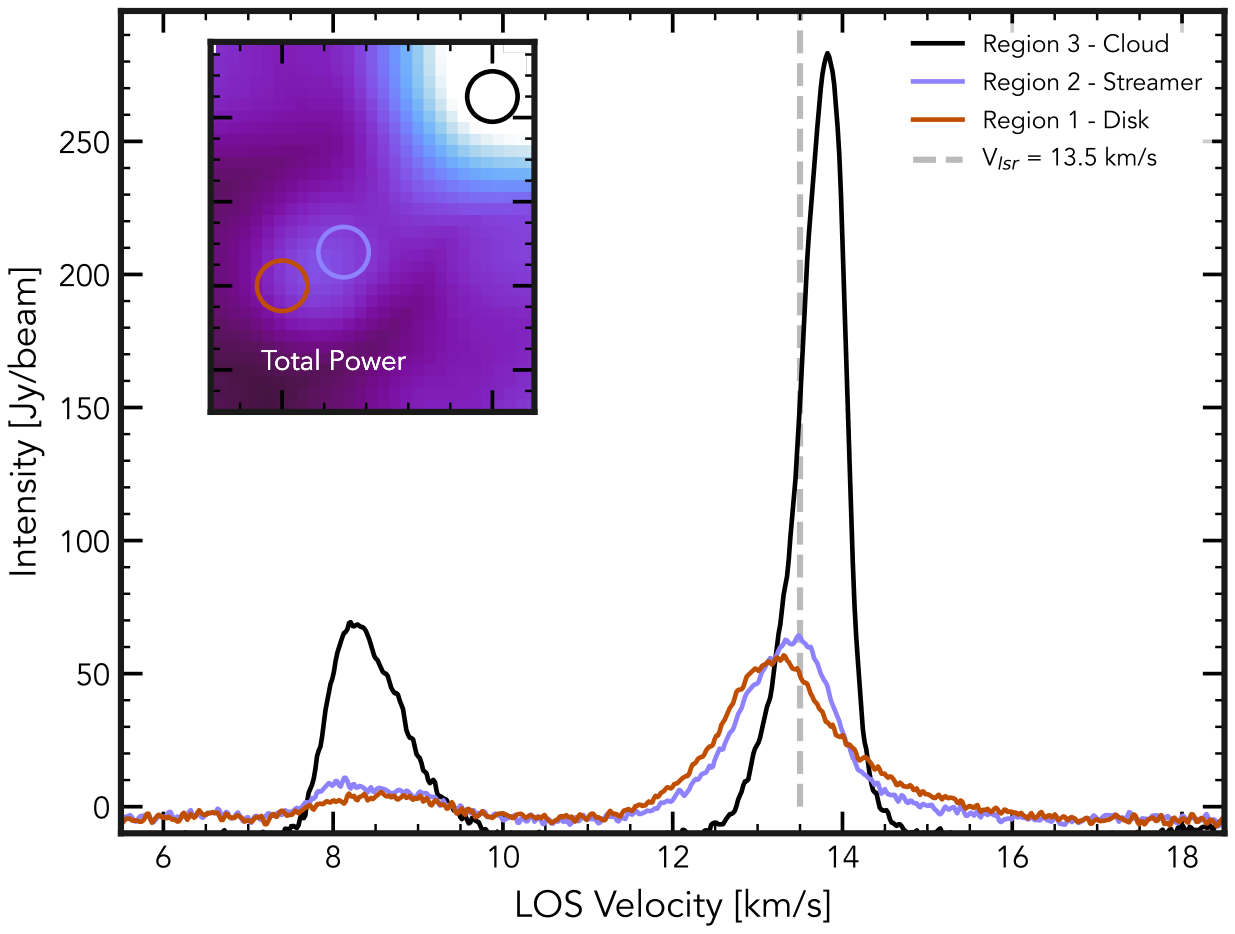

**Figure 11.** Spectra from three points in the $^{12}$CO Total Power data: Region 1 is centered on the disk (orange), Region 2 is centered on the streamer (blue), and Region 3 is centered on the cloud emission, demonstrating that the cloud is kinematically associated with the streamer. The systemic velocity of the disk, 13.5 km s$^{-1}$, is shown as a gray dashed line. The emission at ∼9 km s$^{-1}$ is contamination from the surrounding cloud that the system is embedded in.

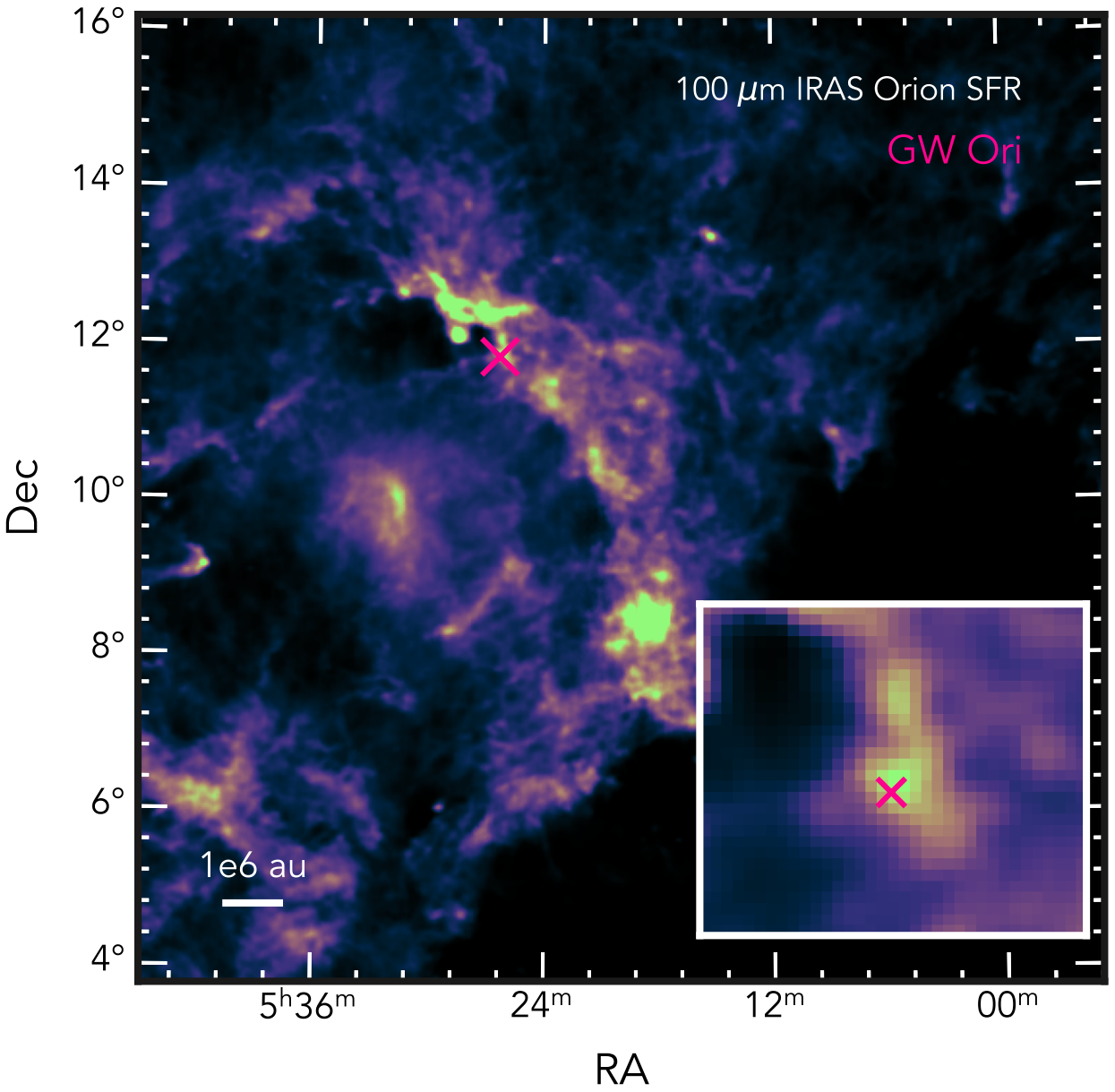

**Figure 12.** IRAS 100 μm image of the Orion star-forming-region (M.-A. Miville-Deschênes & G. Lagache 2005) with the location of GW Ori marked in red. A zoomed-in view of the region around GW Ori is shown in the lower right corner.

where $G$ is the gravitational constant, $M_*$ is the stellar mass, $c_s$ is the sound speed of the parental molecular cloud, and $v_{vel}$ is the relative velocity between the parental molecular cloud and the star (H. Bondi & F. Hoyle 1944). Adopting $M_* = 5.3\,M_\odot$, gas temperature of 15 K, and $v_{rel} \approx 0.5$ km s$^{-1}$ (see Figure 11), we obtain $R_{BH} \leqslant 31{,}000 \pm 18{,}000$ au, where the error on $v_{rel}$ is given as $\sqrt{3}$. The projected distance between GW Ori and the northwest emission in the Total Power image is about 50″ ≈ 28,100 au, well within the BH radius, with the caveat that this is an upper limit, as $v_{rel}$ is simply the RV and does not take into account the unknown proper motion of the cloud. While the true 3D distance cannot be determined, this projected distance supports the hypothesis that BH accretion could be the origin of the streamer. As we showed earlier, the BH framework also predicts a positive correlation between streamer angular momentum and stellar mass that matches our derived value closely (V.-M. Pelkonen et al. 2025), providing independent support for this interpretation.

#### 4.3.2. End Stage of the Infall Event

The properties of the streamer—its low mass, short infall timescale, and mass infall rate roughly an order of magnitude below the stellar accretion rate—suggest that we might be witnessing the final stages of the infall event.

The BH framework allows us to estimate an accretion rate. The BH accretion rate is given by

$$\dot{M}_{BH} = \frac{4\pi G^2 \rho}{(c_s^2 + v_{rel}^2)^{3/2}} M_*^2, \quad (5)$$

where the other variables are the same as in Equation (4) and $\rho$ is the gas density (H. Bondi & F. Hoyle 1944). With the same assumptions as before, and taking $\rho \approx 2.46 m_H 10^3$ cm$^{-3}$ (ambient molecular gas; T. L. Wilson et al. 1999; C. F. McKee & E. C. Ostriker 2007), we find $\dot{M} \approx 2.7 \times 10^{-6}\,M_\odot$ yr$^{-1}$. This is a much higher accretion rate than the one we estimate in Section 3.5 of $3.6 \times 10^{-8}\,M_\odot$ yr$^{-1}$.

The discrepancy between the inferred BH rate and the currently observed infall rate is most naturally explained if the infall is coming to an end. This interpretation is considered by P. Padoan et al. (2025), who estimate a time-dependent accretion rate $\dot{M}_{acc}$ of $1.3 \times 10^{-7}\,M_\odot$ yr$^{-1}$ $(t/1\,\mathrm{Myr})^{-5}(M_{star}/M_\odot)$, where $t$ is the infall time. Our estimated infall time of ∼0.04 Myr likely does not encompass the total infall time if the infall is from BH accretion. Assuming a total infall time of 2 Myr (consistent with simulations in P. Padoan et al. 2025), the estimated mass accretion rate is $1.1 \times 10^{-7}\,M_\odot$ yr$^{-1}$; assuming even larger infall times (e.g., 3 Myr) brings the estimate even lower, converging toward our observed value in GW Ori. Altogether, the observational data consistently points toward the conclusion that the infall event has been winding down, and the streamer we observe today is in its final stage. Future work on the large-scale region surrounding GW Ori and its streamer could help further test the BH scenario and constrain the infall history of the system.

## 5. Conclusions

We have presented new molecular line observations of the streamer in GW Ori, combining data from ALMA programs 2017.1.00286.S and 2022.1.01108.S with multipointing mosaics to capture the full spatial extent of the infall. The streamer is detected in $^{12}$CO and $^{13}$CO, with nondetections in C$^{18}$O and SO. We summarize our main findings below:

1. The streamer extends out to ∼30″, or 12,000 au, in $^{12}$CO when considering the multiresolution Combined Image.
2. Using the optically thin $^{13}$CO emission, we estimate a streamer mass of $1.6 \pm 0.3\ M_{Jup}$, with an upper mass limit of $4.6^{+1.8}_{-0.98}\ M_{Jup}$ measured using C$^{18}$O.
3. We fit the streamer trajectory with TIPSY, finding a clear infall solution that predicts the streamer originates from behind the disk on-sky. The TIPSY fit allowed us to derive properties of the streamer, listed in Table 2.
4. The angular momentum vector of the streamer is most closely aligned with that of the outermost dust ring, with a relative inclination of only 3°, while it is 32° inclined with the angular momentum vector of the innermost dust ring. This tight alignment between the streamer and outermost dust ring angular momentum vectors is strong observational evidence that the streamer is the source of the disk misalignment in GW Ori. The significant misalignment between the outer ring and the inner ring further supports the idea that the outer ring is a second-generation disk assembled from the infalling material.

5. Using the TIPSY fit, we find the specific angular momentum of the streamer to be $1682 \pm 1269$ au km s$^{-1}$ for $^{12}$CO, which is slightly lower than that of the outer disk of GW Ori (2381 au km s$^{-1}$). Currently, the streamer does not have sufficient angular momentum to disrupt the disk, although the range is within error bars.
6. The mass infall rate of $3.6 \times 10^{-8}\,M_\odot\,\mathrm{yr}^{-1}$ is about an order of magnitude lower than the stellar accretion rate. The short infall timescale and the low infall rate indicate that we are witnessing the end stages of the infall event.
7. The Total Power observations reveal a bright source toward the upper left of the FOV, aligned with the trajectory of the streamer. This emission is well within the BH radius, suggesting that BH accretion from the surrounding molecular cloud is the origin of the streamer. The derived streamer angular momentum is furthermore consistent with the BH scaling relation found in numerical simulations.

Our observations establish a direct link between late-stage infall and disk misalignment. We have shown that the streamer in GW Ori has increased the mass available for planet formation, and likely imprinted a distinct angular momentum vector onto the outer disk. These results provide evidence that disks are dynamically connected to their natal star-forming clouds, can help explain the mass budget problem, and provide context for the diversity of orbital architectures observed in exoplanetary systems. The infall also provides observational evidence that BH accretion is responsible for late-stage infall, and links this mechanism to observable disk properties. Future work on shock tracers in GW Ori could help pinpoint where the streamer impacts the disk, and population-level studies of streamers will help determine if these structures are commonplace.

## Acknowledgments

This paper makes use of the following ALMA data: ADS/JAO.ALMA#2017.1.00286.S and ADS/JAO.ALMA#2022.1.01108.S. ALMA is a partnership of ESO (representing its member states), NSF (USA), and NINS (Japan), together with NRC (Canada), NSTC and ASIAA (Taiwan), and KASI (Republic of Korea), in cooperation with the Republic of Chile. The Joint ALMA Observatory is operated by ESO, AUI/NRAO, and NAOJ. The National Radio Astronomy Observatory and Green Bank Observatory are facilities of the U.S. National Science Foundation operated under cooperative agreement by Associated Universities, Inc. Support for this work was provided by the NSF through award SOSPADA-021 from the NRAO. M.B. has received funding from the European Research Council (ERC) under the European Union's Horizon 2020 research and innovation program (PROTOPLANETS, grant agreement No. 101002188). The authors thank UFIT Research Computing for providing computational resources on HiPerGator and support that have contributed to the research results reported in this publication. We thank Andrés F. Izquierdo and Adam Ginsburg for their insight and useful discussion on this project. We thank Eric Villard for helping calibrate the data. We also thank the NRAO and their data scientists for their invaluable help with the calibration and imaging of the datasets.

## Data Availability

The complete set of TIPSY fitting results for $^{12}$CO and $^{13}$CO is available as data behind Figure 4. These files contain the trajectory and subsequent information used in Figures 4, 5, 6, 8, and 9.

## Appendix A
## Additional Plots

Here we present complete channel maps of the streamer in $^{12}$CO and $^{13}$CO (Figures 13 and 14); peak intensity, integrated intensity, and velocity maps of the 7 m data (Figure 15); and an example of the Keplerian masks applied to the data (Figure 16).

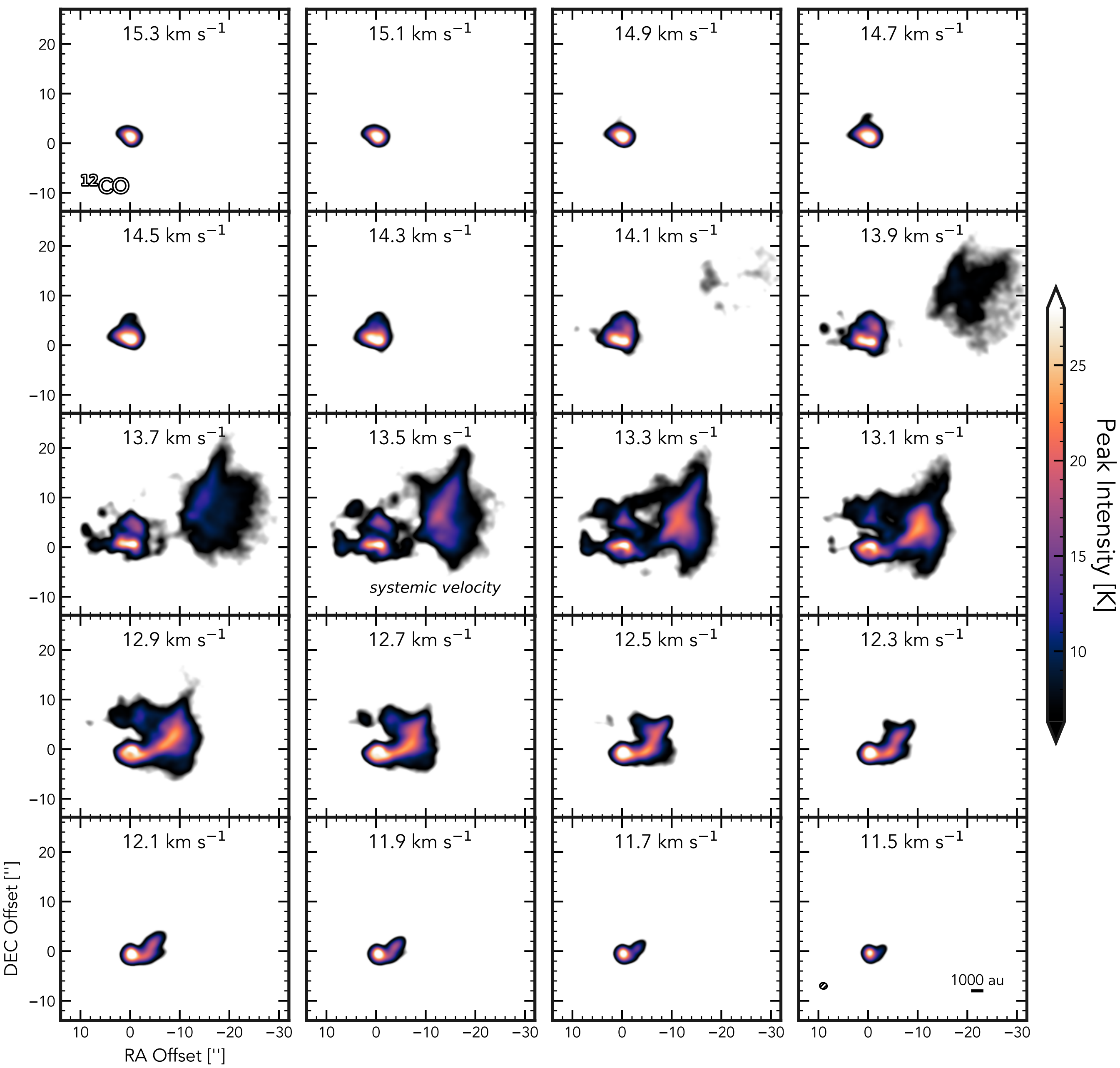

**Figure 13.** Channel maps of $^{12}$CO over the spectral range of the streamer.

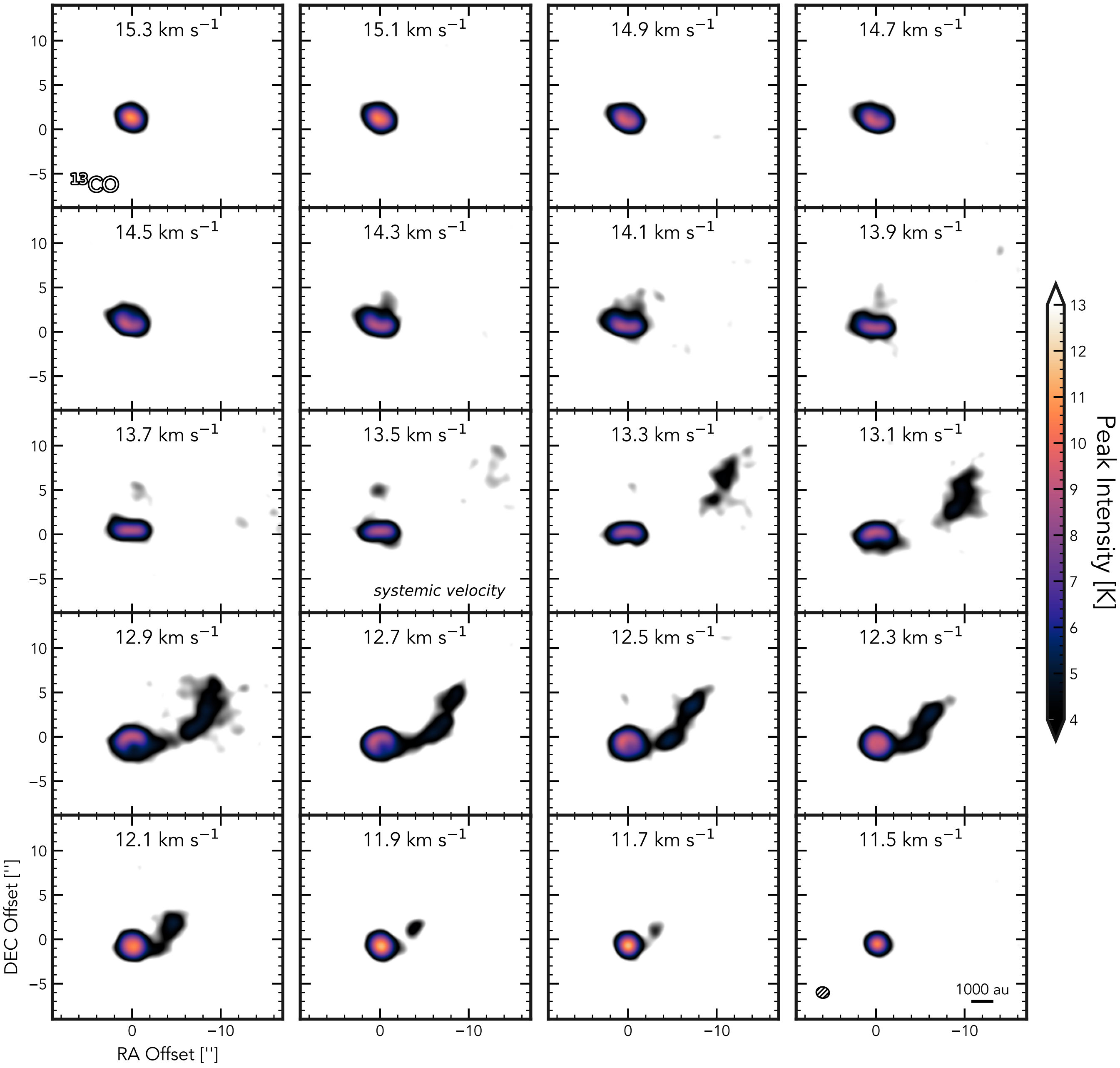

**Figure 14.** Channel maps of $^{13}$CO over the spectral range of the streamer.

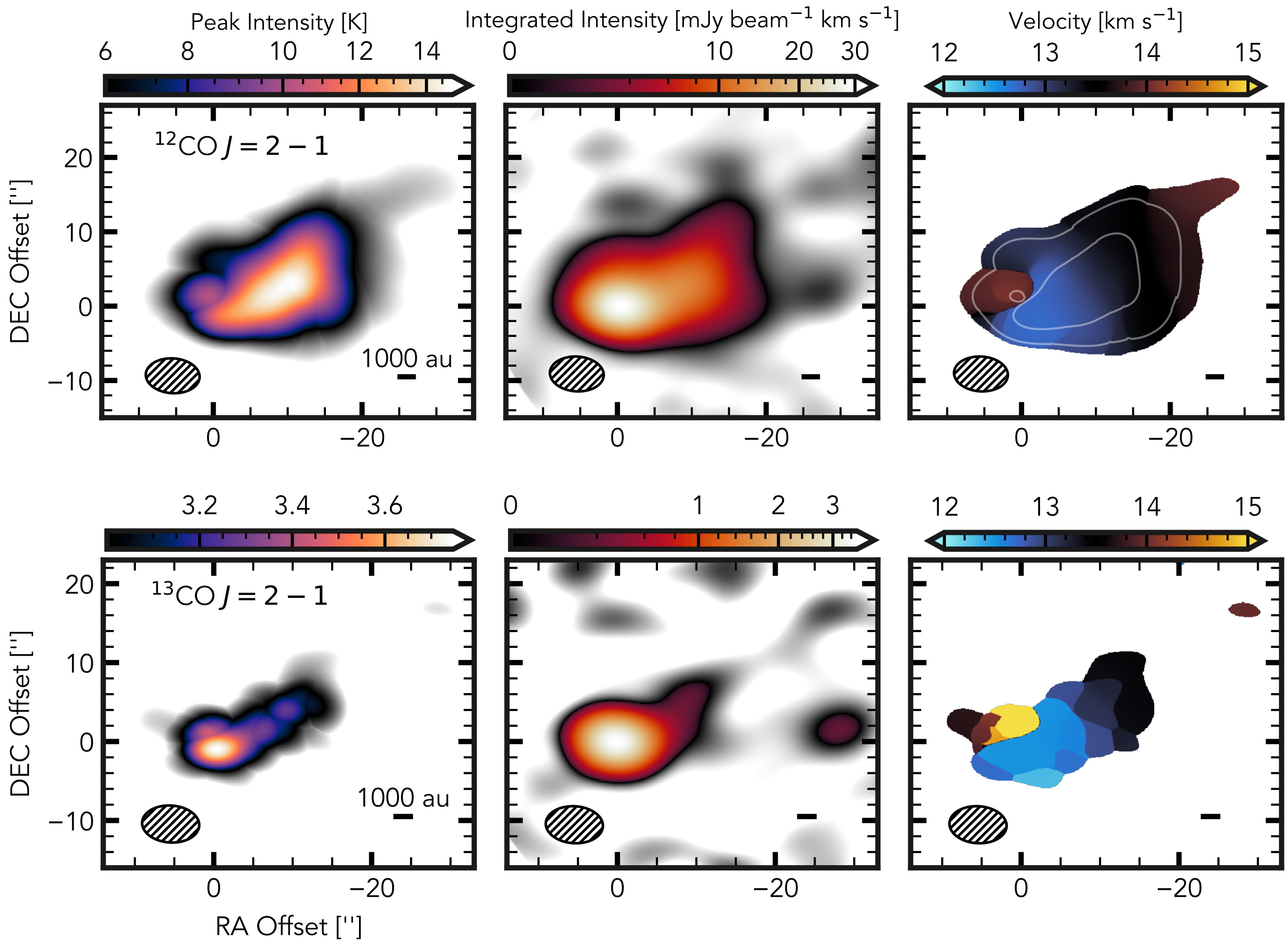

**Figure 15.** Selected moment maps of the Low Resolution Images (7 m center and offset pointings) for $^{12}$CO $J = 2-1$ and $^{13}$CO $J = 2-1$. The emission on the right side of the integrated intensity plots is not directly associated with the disk or streamer, and is likely foreground contamination. It spans 7–8 km s$^{-1}$ (visible in Figure 11), well before the disk and streamer.

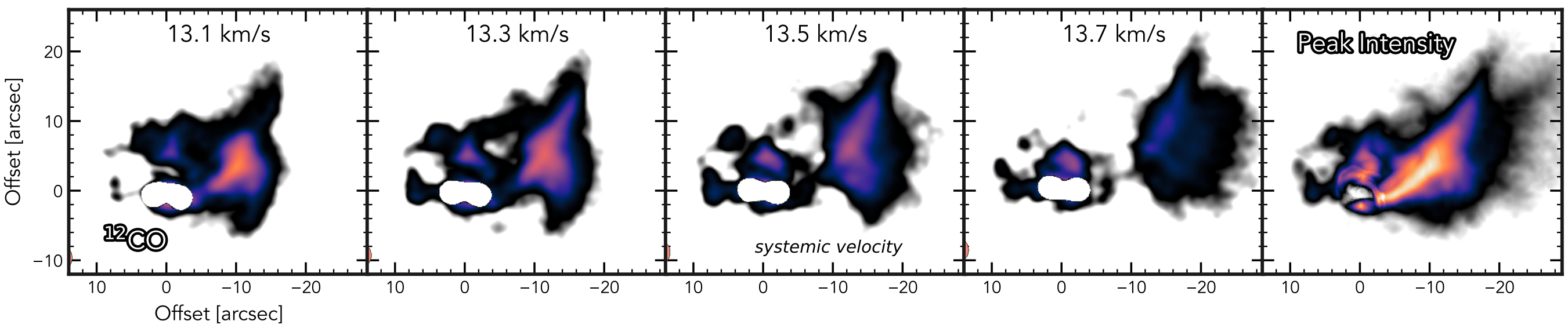

**Figure 16.** Four selected channels of the $^{12}$CO data cube with a Keplerian mask applied; the rightmost panel shows the Keplerian-subtracted peak intensity map. As discussed in Section 3.2, we use GoFish (R. Teague 2019) to create the Keplerian mask, assuming a stellar mass of 5.3 $M_\odot$ and a disk radius of 3.″0 for $^{12}$CO and 2.″0 for $^{13}$CO.

## Appendix B Continuum Flux

While inspecting the data, we found that the continuum flux is approximately 11% higher in the 7 m data than in the 12 m data. To understand the potential cause(s) of this, we investigated the archival continuum 2017 12 m data and the 2022 single-pointing 7 m data in the visibility space. We extracted the continuum windows from the 12 m data (spectral windows 27 and 29), and did the same for the 7 m data (spectral window 24). The 12 m data were subsequently self-calibrated. Self-calibration was attempted on the 7 m data, but we found that with limited integration time that resulted in an SNR of about 10, the self-calibration did not improve the data quality. Figure 17 shows the 12 m and 7 m amplitudes (top), as well as the deprojected and azimuthally averaged real and imaginary visibilities (bottom two panels).

We identify three potential causes of the flux discrepancy. First, the flux calibrator, J0423−0120, has been highly variable, yet its flux density has not been measured sufficiently frequently, particularly around the time of the 12 m observations. Figure 18 shows the Band 6 flux density of J0423−0120 as a function of time, obtained from the ALMA Calibrator Source Catalogue.[6] Since ALMA observed the object first in 2012, its flux density varied between 0.51 Jy and 4.86 Jy. The 12 m observations were carried out on 2018 April 4, and between the two nearest flux measurements in 2018 January and July, there was about a factor of 2.5 increase in flux density. If the calibrator was brighter at the time of observation than the interpolated value used for calibration, the flux of GW Ori's disk would have been systematically underestimated in the 12 m data.

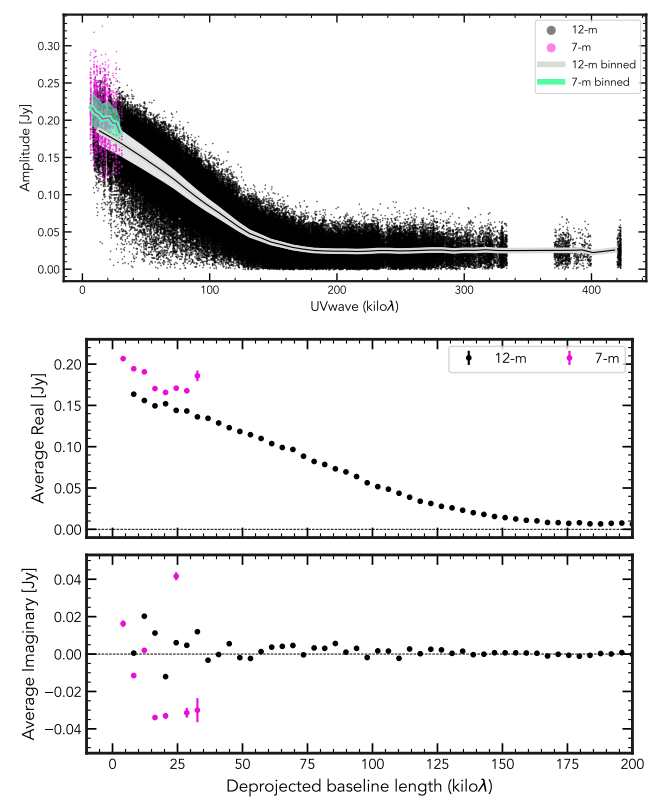

**Figure 17.** Top: amplitude versus UV wave for the 12 m visibilities (black) and the 7 m visibilities (pink). The binned visibilities are shown as a line, and a 10% uncertainty is shown with shaded regions. Bottom: deprojected and azimuthally averaged real and imaginary visibilities.

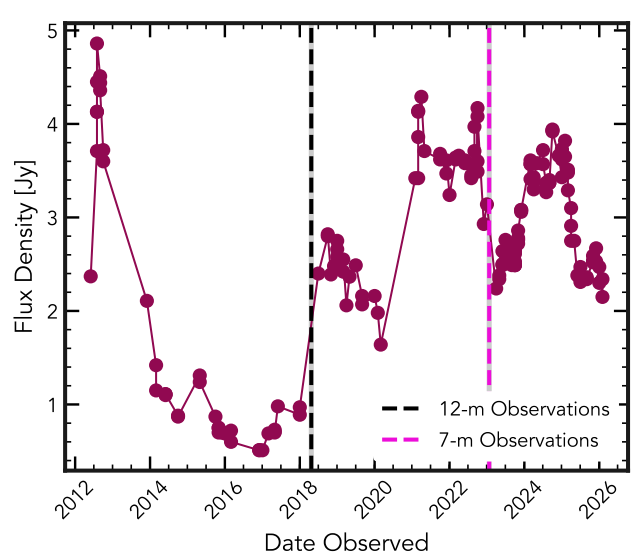

**Figure 18.** Variability of flux calibrator J0423-0120 in Band 6 over time, with the date of the two observations marked with vertical lines.

Second, the 12 m and 7 m arrays inherently sample different regions of the UV plane, with the 7 m array being more sensitive to short baselines and therefore to large-scale, extended emission. Because the GW Ori system contains significant extended emission, the 12 m data may resolve out some of this flux, potentially leading to lower visibility amplitudes.

Finally, a faint continuum source is marginally detected at the 1$\sigma$ level near the western edge of the 7 m array (Figure 19), which is coincident with the emission picked up by the 7 m array seen in the integrated intensity map of $^{12}$CO and $^{13}$CO (see Figure 15). Although this source is nominally cleaned out during imaging as seen in the residual, it may contribute to the higher flux recovered by the 7 m array. We note however that this source is only marginally detected and should be treated with caution. Future deeper observations can confirm the nature of this potential source.

Despite the aforementioned effects, we find that the flux of the line data is broadly consistent between execution blocks and configurations. Still, to account for the combined effect of these uncertainties, we consider a 10% flux uncertainty in our streamer mass estimates in Section 3.2, which is broadly consistent with ALMA's nominal absolute flux calibration uncertainty.

[6] https://almascience.nrao.edu/sc/

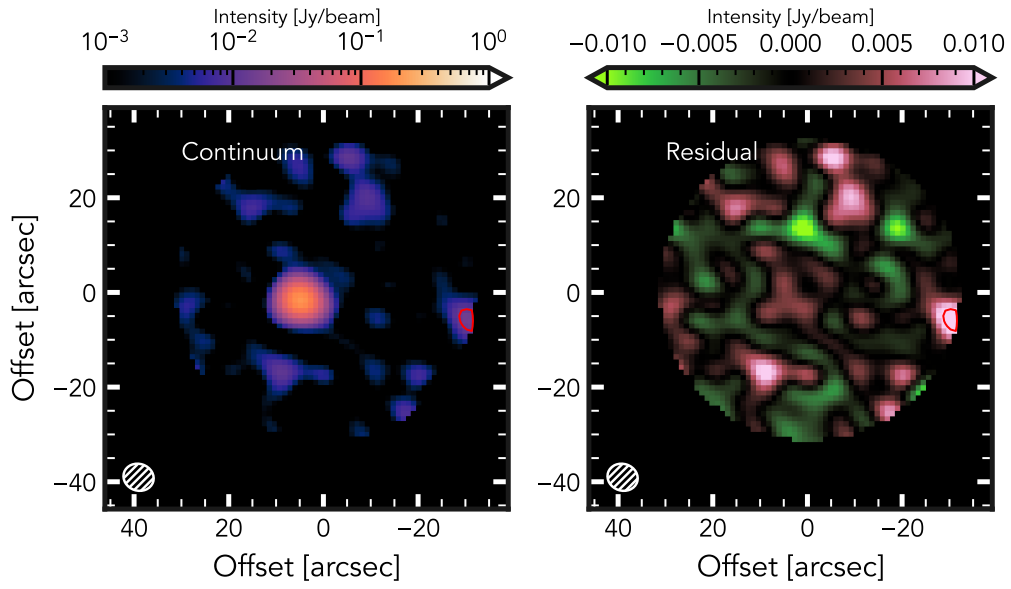

**Figure 19.** Continuum image of the 7 m central pointing (left) with the associated residual after cleaning (right). A faint source at a 1$\sigma$ level of the continuum image can be seen in the far right corner, outlined in red. The beam size is shown in the bottom left corners.

## ORCID iDs

Maria Galloway-Sprietsma https://orcid.org/0000-0002-5503-5476
Jaehan Bae https://orcid.org/0000-0001-7258-770X
Toni Phillips https://orcid.org/0009-0006-2962-9285
Jane Huang https://orcid.org/0000-0001-6947-6072
Myriam Benisty https://orcid.org/0000-0002-7695-7605
Matthew Porter https://orcid.org/0000-0002-9989-4782
Christian Ginski https://orcid.org/0000-0002-4438-1971
Andrew Winter https://orcid.org/0000-0002-7501-9801